# Mathematical modeling of quantum noise and the quality of hardware components of quantum computers

Yu.I. Bogdanov[1], A.Yu. Chernyavskiy[1] , A.S. Holevo[2],

V.F. Luckichev[1], S.A. Nuyanzin[1,3] , A.A. Orlikovsky[1]

[1] Institute of Physics and Technology, Russian Academy of Sciences
[2] Steklov Mathematical Institute, Russian Academy of Sciences
[3] National Research University of Electronic Technology MIET

e-mail: bogdanov@ftian.ru

In the present paper methods and algorithms of modeling quantum operations for quantum computer integrated circuits design are developed. We examine different ways of quantum operation descriptions, including operator-sums, unitary representations, Choi-Jamiolkowski state representations and the corresponding chi-matrices, as well as quantum system evolution operators. The results of modeling of practically important quantum gates: SQiSW (square root of i-SWAP gate), controlled-NOT (CNOT), and controlled Z-transform (CZ) subject to different decoherence mechanisms are presented. These mechanisms include analysis of depolarizing quantum noise and processes of amplitude and phase relaxation. Finally, we consider error correction of phase flip, and the tasks of creating and maintaining the entanglement, as well as its breaking for two- and multi-qubit realizations of quantum operations. Importance of the present analysis for the quality and efficiency of quantum information technologies in practical applications is discussed.

## 1. Introduction

At present dozens of various models of quantum computers are being actively discussed. Among the most prospective and interesting suggestions on quantum registers realization are the projects based on ion traps, nuclear spins, quantum dots, photons, charge, flux and phase states in superconducting structures, atoms in a Rydberg state, quantum states of vacancy centers in diamond, and other [1-4]. The main achievement of the research in the field performed until now is a practical demonstration of validity of physical principles underlying the idea of quantum computations. The main obstacles for building a full-scale quantum computer are a poor development of manufacturing technology for quantum registers relative to requirements, difficulties of measurement and control of quantum states, and of suppression of decoherence caused by quantum noise. Right now, the accuracy of the realization (probability of coincidence between theoretical and experimental quantum states) is only 60-80%, while the required accuracy must be 99.99 % and higher.

One of the most significant bottlenecks in the development of quantum information technologies is the lack of proper methodology for quantum states and processes control. Such a quantum measurements-based methodology is needed to provide an interface between development and implementation of quantum integrated circuits. Mathematically, at the heart of such methodology is the quantitative statistical theory of quantum operations and measurements based on the use of probability-operator valued measures (decompositions of identity) and of completely positive maps of operator algebras in a Hilbert space [5,6]. From the technology viewpoint, methods, algorithms, and computer programs that would be capable to provide an adequate and exhaustive estimation of quality and efficiency of specific quantum information systems must be developed on the basis of the considered mathematical theory [7,8]. Quantum operations are described by reduced dynamics of open quantum systems, based on the concept of complete positivity, as studied in statistical mechanics [9], [10], [11], [12], and in quantum communication theory [13].

It is important to emphasize that the concept of complete positivity can be represented by different means such as extended unitary dynamics of an open quantum system interacting with its environment; Kraus operator decomposition; Choi-Jamilkowski isomorphism; and the formalism of quantum Markov (dynamical) semigroups. We will not limit ourselves to any particular approach, but rather use all of them interconnectively for a comprehensive mathematical modeling of quantum computer integrated circuits. For instance, a unitary representation of quantum operations in extended space is necessary to describe relaxation processes simultaneously with its Hamiltonian evolution (in this case all processes are described by a single Schrödinger's equation). The operator-sum formalism allows one to manifestly decompose a non-unitary evolution of a density matrix into its components as defined by the corresponding Kraus operators. Note, however, that Kraus operators are non-unique (to a wide degree of unitary arbitrariness). The calculation of the Choi – Jamiolkovski chi-matrix allows one to check the fulfillment of the conditions for complete positivity. Such formalism is very important for analyzing the quality of designed gates. In fact, it allows substituting analysis of the evolution of an infinite number of possible states by that of a single state in a higher dimensional Hilbert space. Finally, the concept of the evolution matrix, which simplifies and visualizes the relation between the input and the output density matrices, is very convenient. Moreover, the Choi-Jamiolkowski chi-matrix and the evolution matrix can be easily transformed into one other.

The purpose of our paper is to consistently describe the methods and algorithms of mathematical modeling of quantum operations for the tasks of the quantum computer integrated circuit implementation. We focus on analysis of quantum operations described by the Hamiltonian dynamics of quantum gates that are influenced by environmental quantum noise, in particular, on some practically important gates, such as SQiSW (a square root from i-SWAP), controlled NOT (CNOT), and also controlled Z-transformation (CZ). We take into account various mechanisms of decoherence, including the depolarizing quantum noise, and the processes of amplitude and phase relaxation. Another important goal is the analysis of entanglement dynamics which plays a key role in quantum information technologies. In particular, the analysis of emergence, maintaining and breaking of the entanglement in the two- and multi- qubit operations, subject to influence of quantum noise is important both from fundamental and practical points of view.

The structure of the paper is as follows:

In Section 2, the basics of quantum operations theory are described. We emphasize a definite matrix representation for the main objects of the theory, which is convenient for computer modeling the quantum operations. We consider different ways of describing quantum operations, including operator-sum, unitary representation, the Choi-Jamiolkowski states with corresponding chi-matrixes, and finally, the evolution operator of a quantum system. In Section 3, we discuss the method of modeling amplitude and phase relaxation of qubit states. The method is based on using Kraus operators and unitary representation. One of the complications in the real gates is that the processes of Hamiltonian evolution and relaxation act simultaneously and cannot be separated. In Section 4 we consider modeling of a number of specific gates, including SQiSW, CNOT and CZ, and attempt to solve the practical implementation issues. The problem of quantum error correction in the Choi-Jamiolkowski states framework is also studied in this Section. Section 5 is devoted to the analysis of entanglement for quantum operations in quantum noise environment. We demonstrate that there are two practically useful measures of entanglement for different ways of choosing two subsystems in a Choi-Jamiolkowski state. The final Section 6 summarizes the main conclusions of our work.

## 2. Quantum operations and quantum noise

It is well known that an ideal quantum logical element (gate) performs a unitary transformation of a quantum state (density matrix):

$$\rho_{out} = U \rho_{in} U^{\dagger}. \qquad (1)$$

However, real evolution is never purely unitary. For more realistic models we need to consider the unavoidable interaction of the quantum system with its environment (quantum noise). In the framework of quantum open systems theory, the evolution of states is determined by the operator-sum $\mathbf{E}(\rho) = \sum_k E_k \rho E_k^\dagger$ [5, 6]. Therefore, the relation between the input and the output states is defined by the formula:

$$\rho_{out} = \sum_k E_k \rho_{in} E_k^\dagger , \tag{2}$$

where $E_k$ are the so-called Kraus operators.

The operators $E_k$ in the $s$ -dimensional Hilbert space can be represented by $s \times s$ -matrices. For a unitary transformation there is only one summand defined by the operator $U$ in the sum. Operators $E_k$ must satisfy the constraint to preserve the density matrix trace:

$$Tr(\rho_{out}) = Tr\left( \sum_k E_k \rho_{in} E_k^\dagger \right) = Tr\left( \left( \sum_k E_k^\dagger E_k \right) \rho_{in} \right) = 1 . \tag{3}$$

Here we take into account that the trace is invariant under cyclic permutations. Condition (3) must be met for every density matrix $\rho_{in}$ . This is possible if and only if $E_k$ satisfy the following normalization condition:

$$\sum_k E_k^\dagger E_k = I, \tag{4}$$

where $I$ is the identity matrix of the order $s$.

Non-unitary transformation (2) acting in a Hilbert space of dimension $s$ can be interpreted as a consequence of some unitary transform $U$ in a higher dimensional space. Let us consider $m$ operators $E_k$ ( $k = 1,..., m$ ). Then the $ms \times ms$ dimensional unitary matrix $U$ can be written in the following block form:

$$U = \begin{pmatrix} E_1 & : & : \\ : & : & : \\ E_m & : & : \end{pmatrix}. \tag{5}$$

Here the operators $E_k$ define only the first block-column of the matrix. We can complement the matrix to a unitary by orthogonal complement.

Clearly, the Hermitian conjugate matrix $U^\dagger$ has the form

$$U^+ = \begin{pmatrix} E_1^\dagger & : & E_m^\dagger \\ : & : & : \\ : & : & : \end{pmatrix}. \tag{6}$$

Also, the normalization condition (4) meets the unitary property of $U$ since

$$U^+U = \begin{pmatrix} I & : & : \\ : & : & : \\ : & : & : \end{pmatrix}.$$

The unitary matrix $U$ describes the interaction between an ancillary $m$ -level system (environment) and the considered physical $s$ -level system. States of the environment are enumerated by the columns of the block matrix (5).

The transformation operators $E_k$ can be calculated as matrix elements of the block-matrix $U$

$$E_k = \langle k | U | 0 \rangle, \tag{7}$$

where $|0\rangle$ ,..., $|m\rangle$ are orthonormal states of the environment ($m$-dimensional column vectors)

Note that equation (7) is written in block matrix form which means that $|0\rangle$ stands for $|0\rangle \otimes I_s$, where $I_s$ is the identity matrix of order $s$ and similarly, we should take $\langle k| \otimes I_s$ instead of $\langle k|$, etc.

Unitary representation of a quantum operation assumes that the joint input state of the system and the environment is unentangled and it can be described by the tensor product $|0\rangle\langle 0| \otimes \rho_{in}$. The output state is then $U|0\rangle\langle 0| \otimes \rho_{in} U^{\dagger}$. It can be shown by direct calculation that the reduced density matrix of the last formula is defined by the operator decomposition (2). The joint output state $U|0\rangle\langle 0| \otimes \rho_{in} U^{\dagger}$ of the system and the environment is entangled. Thus possible output environment states describe distinguishable alternatives ($k$-th alternative corresponding to the term $E_k \rho_{in} E_k^{\dagger}$).

An operator-sum representation of a quantum operation guarantees that a Hermitian semidefinite trace-normalized input matrix (i.e. a proper density matrix) will be transformed into a proper output density matrix. It appears that an operator-sum representation guarantees not only positivity, but also the so-called complete positivity of a map (2) [5,6].

Let us explain the physical meaning of the complete positivity condition. Let the considered state $\rho_{in}$ correspond to the system $A$, which is a subsystem of the large system $AB$. Therefore $\rho_{in} = \rho_{in}^{A} = Tr_B\left(\rho_{in}^{AB}\right)$, i.e. the input density matrix $\rho_{in}$ is the reduced density matrix of the joint state $\rho_{in}^{AB}$ obtained by the summation over the degrees of freedom of the subsystem $B$ (subsystem $B$ may be of arbitrary dimension). Let the quantum operation $\mathbf{E}$ act on the subsystem $A$ only, while the subsystem $B$ is unchanged under the identical transformation $\mathbf{I}$. Then the joint system $AB$ is subject to the quantum operation $(\mathbf{E} \otimes \mathbf{I})$. As the result, $\rho_{in}^{AB}$ is transformed into $\rho_{out}^{AB}$ by

$$(\mathbf{E} \otimes \mathbf{I})\left(\rho_{in}^{AB}\right) = \rho_{out}^{AB}. \qquad (8)$$

A natural physical condition is that the output matrix $\rho_{out}^{AB}$ is always positive semidefinite. This condition corresponds to the complete positivity of the map **E**.

The non-triviality of the complete positivity condition is shown by the following example. Taking transpose of a density matrix $\rho_{in}$ leads to some density matrix $\rho_{in}^{T}$, which is generally different from $\rho_{in}$. However, the considered transform it is not completely positive hence it is not a quantum operation. In fact, let us consider a two-qubit density matrix, where the second qubit is ancillary. Then perform a partial transpose operation, when only the first qubit is being transposed. It can be shown that matrix elements are changed by the partial transpose as follows:

$$\rho^{\mathrm{T}^1}_{2k_1+j_2+1,2j_1+k_2+1} = \rho_{2j_1+j_2+1,2k_1+k_2+1}, \qquad (9)$$

where $j_1, j_2, k_1, k_2 = 0,1$.

Let an input state be the maximally entangled Bell-state:

$$|\Phi\rangle = \frac{1}{\sqrt{2}}\left(|00\rangle + |11\rangle\right) = \frac{1}{\sqrt{2}}\begin{pmatrix}1\\0\\0\\1\end{pmatrix}, \qquad \rho_{in} = |\Phi\rangle\langle\Phi| = \frac{1}{2}\begin{pmatrix}1&0&0&1\\0&0&0&0\\0&0&0&0\\1&0&0&1\end{pmatrix}. \qquad (10)$$

The partial transpose of the first qubit leads to:

$$\rho_{out} = \frac{1}{2}\begin{pmatrix}1&0&0&0\\0&0&1&0\\0&1&0&0\\0&0&0&1\end{pmatrix}. \qquad (11)$$

The partial transpose operation can be intuitively represented as follows. Let us divide the $4 \times 4$ density matrix $\rho_{in}$ into four $2 \times 2$ blocks. Then, we can apply the partial transform of the first qubit by the permutation of the upper-right and bottom-left blocks. The partial transpose of the second qubit is the individual transposition of each block.
The eigenvalues of $\rho_{out}$ are equal to $-0.5; 0.5; 0.5; -0.5$. Therefore, the output matrix is not completely positive and is not a proper physical density matrix.
On the basis of the transformation elements, we can easily construct a so-called chi-matrix. This matrix plays a key role in quantum process tomography [3, 14-17]. Let us take the $s \times s$ matrix $E_1$ and rewrite it as column vector $e_1$ of length $s^2$ (we put the second column of $E_1$ under the first one, etc.). The obtained column $e_1$ will be the first column of some matrix $e$. Similarly the matrix $E_2$ will define the second column of $e$, etc.
On the basis of the matrix $e$ we can define the $s^2 \times s^2$ matrix $\chi$:

$$\chi = ee^{\dagger}. \tag{12}$$

It is important that $\chi$ can be regarded as some density matrix in $s^2$-dimensional Hilbert space. Therefore, any quantum operation can be represented by some state in the higher-dimensional Hilbert space. This is called Choi-Jamiolkowski isomorphism [5]. The corresponding state can be described as a joint state of two $s$-dimensional subsystems $A$ and $B$. The trace normalization condition is reflected by the fact that the reduced density matrix $\chi_A$ is equal to the identity $s \times s$ matrix:

$$\chi_A = Tr_B\left(\chi\right) = I. \tag{13}$$

During calculation of different output measurement probabilities, the chi-matrix plays a role totally similar to the role of the density matrix. In fact, let the input state be some pure state $\left|c_{in}\right\rangle$ with the density matrix $\rho_{in} = \left|c_{in}\right\rangle\left\langle c_{in}\right|$. Note that the density matrix $\rho_{in}$ is also a projector as $\rho_{in}^2 = \rho_{in}$. We derive the output state $\rho_{out}$. Let us consider the projection on the pure column vector $\left|c_m\right\rangle$ as a measurement ($\Pi = \left|c_m\right\rangle\left\langle c_m\right|$ is the corresponding projector). Then according to the Born-von Neumann postulate the probability of the considered result is

$$P = tr(\rho_{out}\Pi). \tag{14}$$

Next, let us consider a projective measurement on some equivalent effective state. The effective state is defined by the tensor product of the complex conjugate input state and the output measurement state:

$$\left|\tilde{c}_m\right\rangle = \left|c_{in}^*\right\rangle \otimes \left|c_m\right\rangle. \tag{15}$$

$\tilde{\Pi} = \left|\tilde{c}_m\right\rangle\left\langle\tilde{c}_m\right|$ is a projector that corresponds to this measurement. We can consider chi-matrix as some density matrix and define the probability of the equivalent effective measurement:

$$\tilde{P} = tr(\chi\tilde{\Pi}). \tag{16}$$

It can be proved by direct calculation that these probabilities coincide ($\tilde{P} = P$). Therefore, from the probabilistic point of view a quantum process is fully defined by its chi-matrix or its transformation elements $E_k$.

We may recast the property above in another important form based on the use of an ancillary state (ancilla) and Choi-Jamiolkowski state [5,6,16-19]. Let the considered quantum operation $\mathbf{E}$ act on the s-dimensional system $A$. Let us add an ancillary $s$-dimensional system $B$ and consider the joint system $AB$. Then let us input the maximally entangled state

$$|\Phi\rangle = \frac{1}{\sqrt{s}}\sum_{j=1}^{s} |j\rangle \otimes |j\rangle \qquad (17)$$

to our operation. Here the first factor in tensor product corresponds to the subsystem A, and the second corresponds to the subsystem B.
The output state of such process is called the Choi-Jamiolkowski state. Let the identical transform $\mathbf{I}$ act on the subsystem B. Then the transform $(\mathbf{I} \otimes \mathbf{E})$ acts on the entire system $AB$.
It appears that if the density matrix $(|\Phi\rangle\langle\Phi|)$ is submitted to the input, then the chi-matrix is obtained at the output. However, the trace of this matrix is equal to 1, i.e.

$$(\mathbf{I} \otimes \mathbf{E})(|\Phi\rangle\langle\Phi|) = \rho_\chi \text{, где } \rho_\chi = \frac{1}{s}\chi. \qquad (18)$$

The validity of the result can be checked directly:

$$\rho_\chi = \frac{1}{s}\sum_{j, j_1, k} |j\rangle\langle j_1| \otimes E_k |j\rangle\langle j_1| E_k^\dagger = \frac{1}{s}\chi \qquad (19)$$

We illustrate the above considerations by the following figure.

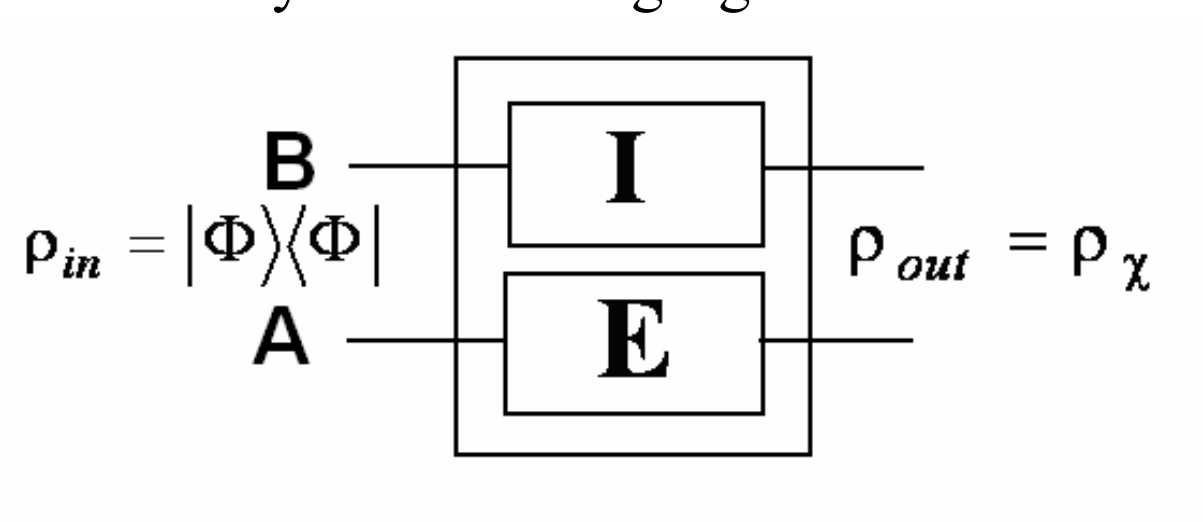

**Fig. 1.** Quantum scheme of Choi-Jamiolkowski state calculation.

Previously, we considered how to construct a chi-matrix by transformation elements $E_k$. The solution of the inverse problem is as well straightforward. We need to diagonalize the chi-matrix:

$$\chi = U_\chi D_\chi U_\chi^\dagger. \qquad (20)$$

Here $D_\chi$ is a diagonal matrix with the eigenvalues of $\chi$ on the diagonal. All these eigenvalues are non-negative since the matrix $\chi$ is positive semidefinite. We arrange the eigenvalues in non-increasing order. The columns of $U_\chi$ are the eigenvectors of $\chi$. Then the matrix $e$ can be calculated by the following equation:

$$e = U_\chi D_\chi^{1/2}. \qquad (21)$$

The number of nonzero eigenvalues of the chi-matrix is called the rank $r$ of the quantum operation. It is obvious that $1 \le r \le s^2$. Thus, an arbitrary quantum operation can be reduced to the form consisting of no more than $s^2$ matrices $E_k$. The case of $r = 1$ corresponds to a unitary transformation. Matrix $D_\chi$ can be reduced to the size $r \times r$ by eliminating the zero columns and rows. Respectively, the matrix $U_\chi$ must be reduced to $s^2 \times r$ size by keeping only the first $r$ columns. Then the matrix $e$ will be of the size $s^2 \times r$, and the equations $\chi = U_\chi D_\chi U_\chi^\dagger = ee^\dagger$ will remain correct. Let us stress that the matrix $e$ and, correspondingly, matrices $E_k$ are defined in non-unique way. Let the matrix $e$ consist of $m$ columns (there are $m$ transformation elements $E_k$, $k = 1,2,...,m$ in the operator-sum) and have the dimension equal to $s^2 \times m$. Clearly, the chi-matrix is invariant under the following transformation:

$$e \to e' = eU\,, \qquad (22)$$

where $U$ is an $m \times m$ unitary matrix.

The new matrices $E'_k$, which are unitarily equivalent to the initial matrices $E_k$, correspond to the new matrix $e'$.

Note that we can make $m = r$ by optimizing the number of transformation elements. The initial number of transformation elements $m$ can be even greater than $s^2$ (in general, it can be arbitrarily large). However, the response of a quantum system can always be defined by just $s^2$ transformation operators, no matter how many elementary "noise" operators are influencing this system. This is an important informational confinement property of finite-dimensional quantum systems.

A chi-matrix can be specified in different representations defined by sets of basis matrices. In all the discussion above we used the so-called standard representation. Let us define this representation explicitly. Let $|j\rangle$ be a ket-vector (column), whose $j$-th element is equal to 1 and the others are zeros and let $\langle k|$ be the corresponding bra-. Consider the matrix unit $|j\rangle\langle k|$, whose single nonzero element (unit) is at the intersection of $j$-th row and $k$- th column. Let indices $j$ and $k$ take values from 1 to $s^2$ ($j,k = 1,2,...,s^2$). In total, there are $s^4$ such matrices. It is obvious that the chi-matrix can be written in the following form:

$$\chi = \sum_{j,k} \chi_{jk} |j\rangle\langle k|\,. \qquad (23).$$

Here the set of $s^4$ matrices $|j\rangle\langle k|$ form the basis. The decomposition coefficients are obviously different for a transition from the matrices $|j\rangle\langle k|$ to another basis sets. This corresponds to different chi-matrix representations.

For example, let us describe a transition from the standard basis described above to a representation defined by the Pauli matrices [3]. Consider a one-qubit basis set defined by the following four $2 \times 2$ matrices

$$I = E/\sqrt{2}\,, X = \sigma_x/\sqrt{2}\,, Y = -i\sigma_y/\sqrt{2}\,, Z = \sigma_z/\sqrt{2}\,. \qquad (24)$$

Then a two-qubit basis set will consist of all pairwise tensor products of the considered matrices (16 matrices), a three-qubit set - of all tensor triplets (64 matrices) etc. We limit our consideration to the case of two-qubit quantum operations (generalizations to other cases is straightforward).

A transition to the two-qubit basis set is performed by calculation of all pairwise tensor products of one-qubit matrices:

$II = I \otimes I$ ; $IX = I \otimes X$ ; $IY = I \otimes Y$ ; $IZ = I \otimes Z$ ;
$XI = X \otimes I$ ; $XX = X \otimes X$ ; $XY = X \otimes Y$ ; $XZ = X \otimes Z$ ;
$YI = Y \otimes I$ ; $YX = Y \otimes X$ ; $YY = Y \otimes Y$ ; $YZ = Y \otimes Z$ ;
$ZI = Z \otimes I$ ; $ZX = Z \otimes X$ ; $ZY = Z \otimes Y$ ; $ZZ = Z \otimes Z$ . (25)

We obtain 16 columns by stretching vertically all 16 matrices. For example, to obtain column $e_{II}$ we need to write the second column of the matrix $II$ under the first one, etc. As the result, we get 16 columns that generate the required $16 \times 16$ unitary transition matrix $U_0$ as follows:

$$U_0 = [e_{II}\ \ e_{IX}\ \ e_{IY}\ \ e_{IZ}\ \ e_{XI}\ \ e_{XX}\ \ e_{XY}\ \ e_{XZ}\ \ e_{YI}\ \ e_{YX}\ \ e_{YY}\ \ e_{YZ}\ \ e_{ZI}\ \ e_{ZX}\ \ e_{ZY}\ \ e_{ZZ}]. \qquad (26)$$

Finally, the transition from the matrix $\chi$ in the standard basis to the matrix $\chi'$ in the basis defined by the Pauli matrices is specified by the following unitary transform:

$$\chi' = U_0^{\dagger} \chi U_0. \qquad (27)$$

Transition to any other basis can be described in a similar manner.

The sets of the basis matrices considered above are orthonormal. Generally, a set of basis matrices $a_j$ ( $j = 1,...,m$ ) is orthonormal if

$$Tr(a_j a_k^\dagger) = \delta_{jk}, \qquad j,k = 1,...,m. \tag{28}$$

Let us now consider an important connection between the chi-matrix and the evolution operator of a quantum state.

Let density matrices $\rho_{in}$ and $\rho_{out}$ be stretched into $s^2$-length columns. Then, by virtue of linearity of quantum operation, we can write:

$$\rho_{out} = G\rho_{in}. \tag{29}$$

Here $G$ specifies the corresponding $s^2 \times s^2$ evolution matrix.

It is important to note that matrices $G$ and $\chi$ consist of the same elements (permuted).

It can be shown that the correspondence between the indices of these matrices is given by the following relation:

$$G_{m+s(n-1),j+s(k-1)} = \chi_{(j-1)s+m,(k-1)s+n} \qquad j,k,m,n = 1,2,..,s \tag{30}$$

Time-homogenous Markovian processes are an important special case of such correspondence. Let $G(\Delta t)$ be an evolution matrix for the time interval $\Delta t$. Then $N$ steps of the evolution (i.e. the evolution during time $t = N\Delta t$ ) are defined by:

$$G(t) = (G(\Delta t))^N \tag{31}$$

A Markovian condition of a quantum operation means that the environment "remembers" its correlations with the system for an infinitesimally small period of time. In the language of a unitary representation of a quantum operation this can be explained by the following consideration. As usual, consider an unentangled joint input state of a system and its environment that is given by the tensor product $|0\rangle\langle 0| \otimes \rho_{in}$. We obtain an entangled joint state $U|0\rangle\langle 0| \otimes \rho_{in}U^\dagger$ as the effect of unitary evolution $U$. The process of "forgetting" the correlations between the system and its environment is characterized by the fact that after short time $\Delta t$ the joint state becomes unentangled again. This state is given by the tensor product $|0\rangle\langle 0| \otimes \rho_{out}$, i.e.:

$$U|0\rangle\langle 0| \otimes \rho_{in}U^\dagger \xrightarrow{\text{decoherence}} |0\rangle\langle 0| \otimes \rho_{out}. \tag{32}$$

It is very useful to be able to separate "purely quantum" noise while practically working with noisy quantum gates. In order to do so we need to represent the considered quantum operation $\mathbf{E}$ in the form:

$$\mathbf{E} = \mathbf{E}\mathbf{E}_0^{-1}\mathbf{E}_0 = \tilde{\mathbf{E}}\mathbf{E}_0. \tag{33}$$

Here $\mathbf{E}_0$ is the ideal unitary quantum operation, $\mathbf{E}_0^{-1}$ is the inverse operation to $\mathbf{E}_0$. We also defined the new quantum operation $\tilde{\mathbf{E}} = \mathbf{E}\mathbf{E}_0^{-1}$, which effectively corresponds to the quantum noise in the gate [3]. The operation $\tilde{\mathbf{E}}$ allows one to estimate the quality of a quantum gate (the closer the operation to the identity transformation, the higher is the quality of the operation).

We denote the chi-matrix that corresponds to the operation $\tilde{\mathbf{E}}$ by $\tilde{\chi}$ [1]

[1] This matrix is quite useful in problems of quantum processes' tomography (A.N. Korotkov- private communication)

## 3. Amplitude and phase relaxation of qubit states

Let us start with a simple but a very important case of pure dephasing.

The Kraus operators of pure dephasing are given by

$$E_0 = \begin{pmatrix} 1 & 0 \\ 0 & \sqrt{1-\gamma} \end{pmatrix}, \qquad E_1 = \begin{pmatrix} 0 & 0 \\ 0 & \sqrt{\gamma} \end{pmatrix}. \tag{34}$$

Using (2) we can easily calculate transformation of a density matrix

$$\rho_{out} = E_0 \rho_{in} E_0^{\dagger} + E_1 \rho_{in} E_1^{\dagger}. \tag{35}$$

Let us parameterize the input state as follows:

$$\rho_{in} = \begin{pmatrix} a & b \\ b^{*} & 1-a \end{pmatrix}. \tag{36}$$

Then

$$\rho_{out} = \begin{pmatrix} a & b\sqrt{1-\gamma} \\ b^{*}\sqrt{1-\gamma} & 1-a \end{pmatrix}. \tag{37}$$

We see that diagonal elements of the density matrix are invariant for the pure dephasing case. These elements determine the populations of the energy levels.

Historically, a spin-spin nuclear relaxation was one of the first phenomenon's used for studies of quantum processes [20]. Nuclear spins and other physical systems experiments show that non-diagonal elements characterizing coherence usually decrease exponentially with time. Therefore it is natural to use the following parameterization:

$$\sqrt{1-\gamma} = \exp\left(-\frac{t}{T_2^{pure}}\right), \quad \gamma = 1 - \exp\left(-\frac{2t}{T_2^{pure}}\right). \tag{38}$$

The time parameter $T_2^{pure}$ introduced here characterizes pure dephasing.

Finally, we have the following equation for the pure dephasing process:

$$\rho_{out} = \begin{pmatrix} a & b\exp\left(-\frac{t}{T_2^{pure}}\right) \\ b^{*}\exp\left(-\frac{t}{T_2^{pure}}\right) & 1-a \end{pmatrix} \tag{39}$$

Let us note the correspondence between the pure dephasing process and phase flip. A phase flip (Z-error) is described by the following Kraus operators:

$$E_0 = \sqrt{1-p}\begin{pmatrix} 1 & 0 \\ 0 & 1 \end{pmatrix}, \quad E_1 = \sqrt{p}\begin{pmatrix} 1 & 0 \\ 0 & -1 \end{pmatrix}, \tag{40}$$

where $p$ is the probability of error.

It is easy to see that the two considered processes coincide if

$$1 - 2p = \sqrt{1-\gamma}, \qquad p \le \frac{1}{2}. \tag{41}$$

Consequently, we obtain the correspondence between the error probability and the pure dephasing time:

$$p = \frac{1 - \exp\left(-\frac{t}{T_2^{pure}}\right)}{2}, \quad T_2^{pure} = -\frac{t}{\ln(1-2p)}. \tag{42}$$

Initially $(t=0)$ $p$ is equal to zero (no decoherence), and we have full decoherence ($p \to 1/2$) as $t \to \infty$ .

The so-called amplitude relaxation is the next important process we consider. In this case the Kraus operators are given by:

$$E_0 = \begin{pmatrix} 1 & 0 \\ 0 & \sqrt{1-\gamma} \end{pmatrix}, \quad E_1 = \begin{pmatrix} 0 & \sqrt{\gamma} \\ 0 & 0 \end{pmatrix}. \qquad (43)$$

The parameter $\gamma$ defines the relaxation (leap) probability. This leap is related to the transfer from the excited state $|1\rangle$ to the ground state. We assume that $|0\rangle$ is the final state after the amplitude relaxation.

Once again let $\rho_{in} = \begin{pmatrix} a & b \\ b^* & 1-a \end{pmatrix}$ , then

$$\rho_{out} = \begin{pmatrix} a+\gamma(1-a) & b\sqrt{1-\gamma} \\ b^*\sqrt{1-\gamma} & (1-\gamma)(1-a) \end{pmatrix}. \qquad (44)$$

We see that the amplitude relaxation affects both diagonal and non-diagonal elements. Note that generally it is impossible to provide an upper level population damping without the loss of coherence, due to the loss of density matrix positive semidefiniteness property.
Historically, a spin-lattice nuclear relaxation was one of the most important systems in the studies of the amplitude relaxation process [20].
Let us introduce the parameter $T_1$ that characterizes time-exponential relaxation of the upper level

$$\gamma = 1 - \exp\left(-\frac{t}{T_1}\right), \quad \sqrt{1-\gamma} = \exp\left(-\frac{t}{2T_1}\right). \qquad (45)$$

and discuss the difference between these equations and formulas for the pure dephasing. Time $T_1$ parameterizes the speed of relaxation of the diagonal elements of the density matrix. For non-diagonal elements the relaxation is two times slower. Thus

$$\rho_{out} = \begin{pmatrix} 1-(1-a)\exp\left(-\frac{t}{T_1}\right) & b\exp\left(-\frac{t}{2T_1}\right) \\ b^*\exp\left(-\frac{t}{2T_1}\right) & (1-a)\exp\left(-\frac{t}{T_1}\right) \end{pmatrix}. \qquad (46)$$

One can easily see that the effect of both processes applied simultaneously is the following:

$$\rho_{out} = \begin{pmatrix} 1-(1-a)\exp\left(-\frac{t}{T_1}\right) & b\exp\left(-\frac{t}{T_2}\right) \\ b^*\exp\left(-\frac{t}{T_2}\right) & (1-a)\exp\left(-\frac{t}{T_1}\right) \end{pmatrix}. \qquad (47)$$

Here we introduced the new parameter $T_2$ such that

$$\frac{1}{T_2} = \frac{1}{T_2^{pure}} + \frac{1}{2T_1}. \qquad (48)$$

Therefore,

$$T_2 \le 2T_1. \qquad (49)$$

Through the experimental measurement of a density matrix, we can find the parameters $T_1$ and $T_2$ . Then, we can easily calculate the pure dephasing parameter:

$$T_2^{pure} = \frac{2T_1T_2}{2T_1 - T_2}. \tag{50}$$

Next, we shall consider a unitary representation of amplitude and phase relaxation processes. This representation will be used in Sections 4 and 5 for analysis of implementation of the algorithms modeling the quantum gates. Consider the phase relaxation first; the Hamiltonian that describes the interaction between a physical qubit and a ancillary qubit that simulates the environment is

$$H = i\left(b^\dagger \otimes a - b \otimes a^\dagger\right). \tag{51}$$

Here $a$ , $a^\dagger$ , $b$ , $b^\dagger$ are the annihilation and creation operators of the main and ancillary qubits respectively, $a = \begin{pmatrix} 0 & 1 \\ 0 & 0 \end{pmatrix}$, $a^\dagger = \begin{pmatrix} 0 & 0 \\ 1 & 0 \end{pmatrix}$, and similarly for *b*.

The unitary evolution defined by this Hamiltonian is

$$U = \exp\left(-i\, H\ \theta\right) = \exp\left(\left(b^\dagger \otimes a - b \otimes a^\dagger\right)\ \theta\right). \tag{52}$$

Let $A = \left(b^\dagger \otimes a - b \otimes a^\dagger\right) = \begin{pmatrix} 0 & -a^\dagger \\ a & 0 \end{pmatrix}$.

Then $A^2 = -B$ , where $B = \begin{pmatrix} P_1 & 0 \\ 0 & P_0 \end{pmatrix}$, $P_0$ and $P_1$ are the projectors:

$$P_0 = \begin{pmatrix} 1 & 0 \\ 0 & 0 \end{pmatrix},\ P_1 = \begin{pmatrix} 0 & 0 \\ 0 & 1 \end{pmatrix}.$$

The next powers of the matrix *A* are $A^3 = -A$ , $A^4 = B$ etc.

It can be shown that in Taylor's series expansion the odd powers will yield sine and even powers cosine. As the result we obtain:

$$U = \begin{pmatrix} P_0 + P_1 \cos\theta & -a^\dagger \sin\theta \\ a \sin\theta & P_1 + P_0 \cos\theta \end{pmatrix}. \tag{53}$$

Summation over degrees of freedom of the environment gives the following Kraus operators:

$$E_0 = \langle 0|U|0\rangle = P_0 + P_1 \cos\theta = \begin{pmatrix} 1 & 0 \\ 0 & \cos\theta \end{pmatrix}, \tag{54}$$

$$E_1 = \langle 1|U|0\rangle = a \sin\theta = \begin{pmatrix} 0 & \sin\theta \\ 0 & 0 \end{pmatrix}. \tag{55}$$

Thus, the relationship between the parameters $\theta$ and $\gamma$ is given by the following formula:

$$\gamma = \sin^2\theta. \tag{56}$$

From equations (45) and (56), we get the relationship between the fictitious time $\theta$ and the real time $t$

$$\theta = \arcsin\left(\sqrt{1 - \exp\left(-\frac{t}{T_1}\right)}\right), \tag{57}$$

$$t = -T_1 \ln\left(\cos^2\theta\right). \tag{58}$$

This relationship is illustrated in the following figure.

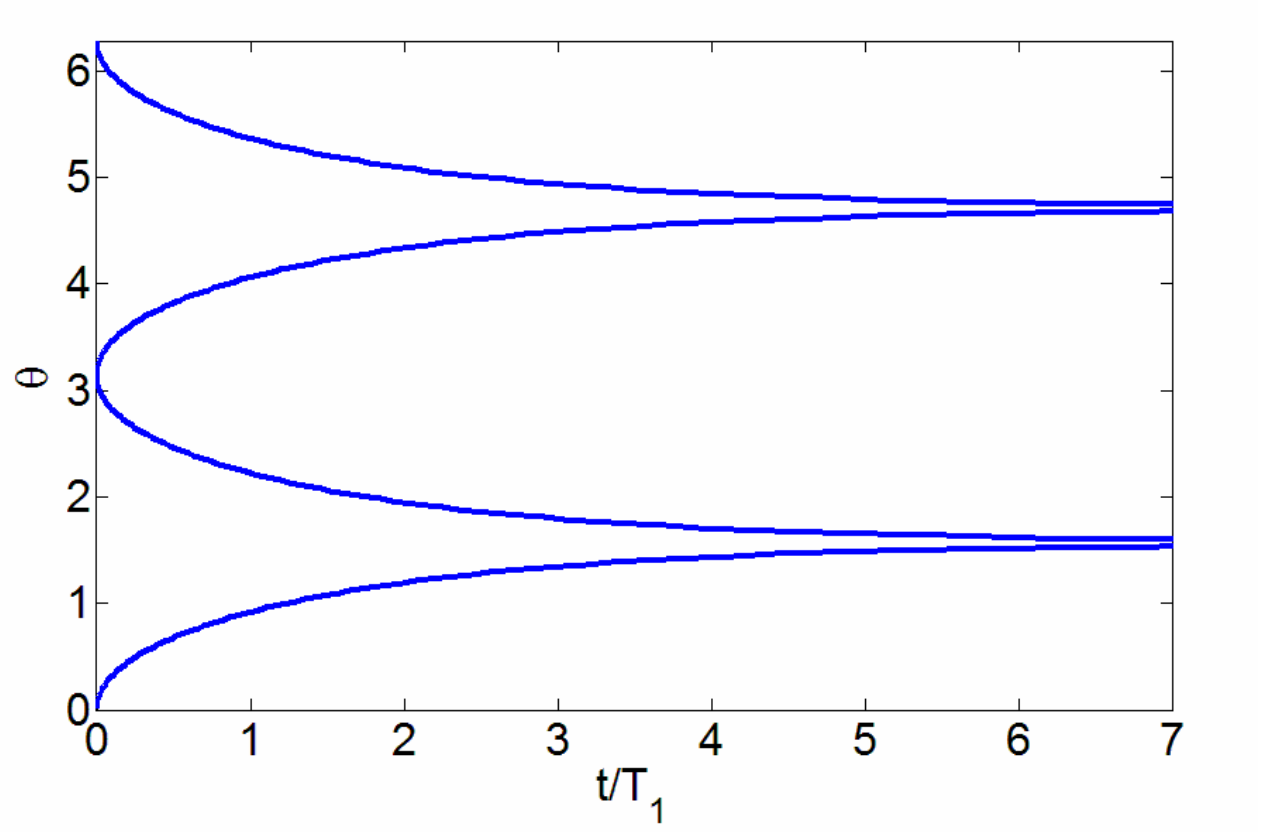

**Fig. 2.** The relationship between the real and fictitious times in the problem of the amplitude relaxation unitary representation.

We will provide another useful interpretation of this result. If the initial state of the environment is $|0\rangle$ and the state of the physical qubit is $\alpha|0\rangle+\beta|1\rangle$, then the unitary evolution of the joint system is:

$$U|0\rangle\left(\alpha|0\rangle+\beta|1\rangle\right)=\alpha|00\rangle+\beta\left(\cos\theta|01\rangle+\sin\theta|10\rangle\right). \qquad (59)$$

We see that two distinguishable alternatives appear as the result of interaction of the qubit with its environment. If we obtain $|0\rangle$ after the measurement of the environment then

$$\begin{pmatrix}\alpha\\ \beta\end{pmatrix}\to\begin{pmatrix}\alpha\\ \beta\cos\theta\end{pmatrix}. \qquad (60)$$

This is the $E_0$ process since

$$E_0\begin{pmatrix}\alpha\\ \beta\end{pmatrix}=\begin{pmatrix}\alpha\\ \beta\cos\theta\end{pmatrix}. \qquad (61)$$

If we obtain $|1\rangle$ after the measurement of the environment then

$$\begin{pmatrix}\alpha\\ \beta\end{pmatrix}\to\begin{pmatrix}\beta\sin\theta\\ 0\end{pmatrix}. \qquad (62)$$

This is the $E_1$ process, because

$$E_1\begin{pmatrix}\alpha\\ \beta\end{pmatrix}=\begin{pmatrix}\beta\sin\theta\\ 0\end{pmatrix}. \qquad (63)$$

Next, let us consider the interaction Hamiltonian for the phase relaxation modeling

$$H=\left(b+b^{\dagger}\right)\otimes a^{\dagger}a. \qquad (64)$$

Here the operators $b$ and $b^{\dagger}$ correspond to the environment qubits, while $a$ and $a^{\dagger}$ correspond to the physical qubits.

The unitary evolution defined by this Hamiltonian is

$$U=\exp\left(-i\,H\,\theta\right)=\exp\left(-i\,\theta\left(b+b^{\dagger}\right)\otimes a^{\dagger}a\right). \qquad (65)$$

Let us represent the Hamiltonian and the corresponding unitary transform in the explicit matrix form:

$$a^{+}a = \begin{pmatrix} 0 & 0 \\ 1 & 0 \end{pmatrix}\begin{pmatrix} 0 & 1 \\ 0 & 0 \end{pmatrix} = \begin{pmatrix} 0 & 0 \\ 0 & 1 \end{pmatrix} = P_1,$$

$$b + b^{+} = \begin{pmatrix} 0 & 1 \\ 1 & 0 \end{pmatrix} = \sigma_x,$$

$$H = \left(b + b^{\dagger}\right) \otimes a^{\dagger}a = \begin{pmatrix} 0 & P_1 \\ P_1 & 0 \end{pmatrix},$$

the square of this matrix being

$$H^2 = \begin{pmatrix} P_1 & 0 \\ 0 & P_1 \end{pmatrix}.$$

Expansion of $U$ into Taylor's series gives

$$U = \begin{pmatrix} P_0 + P_1\cos\theta & iP_1\sin\theta \\ iP_1\sin\theta & P_0 + P_1\cos\theta \end{pmatrix}. \tag{66}$$

Summation over degrees of freedom of the environment gives

$$E_0 = \langle 0|U|0\rangle = P_0 + P_1\cos\theta = \begin{pmatrix} 1 & 0 \\ 0 & \cos\theta \end{pmatrix}, \tag{67}$$

$$E_1 = \langle 1|U|0\rangle = iP_1\sin\theta = i\begin{pmatrix} 0 & 0 \\ 0 & \sin\theta \end{pmatrix}. \tag{68}$$

Note that the phase coefficient $i$ is insignificant in the last equation, because it does not affect the Kraus decomposition.

Therefore the relationship between parameters $\theta$ and $\gamma$ is again

$$\gamma = \sin^2\theta. \tag{69}$$

The relation between the fictitious time $\theta$ and the real time $t$ is determined by the following formula

$$\theta = \arcsin\left(\sqrt{1 - \exp\left(-\frac{2t}{T_2^{pure}}\right)}\right). \tag{70}$$

This relationship is analogous to (57).

**4. Mathematical modeling of quantum operations**

The approach in this paper is based on the use of the Choi-Jamiolkowski states and of the unitary representation of open quantum system. A quantum operation that describes the action of a noisy quantum gate can be given via chi-matrix, a G-matrix (the evolution matrix), or via the Kraus operators. The goal of the modeling is to compute these matrices.

As an example, let us consider the method of two-qubit gate modeling subject to phase and amplitude relaxations. In this case, we can use 6 ancillary qubits along with the two physical qubits. Qubits 1 and 2 represent ancillas of the relative Choi-Jamiolkowski state. These two qubits together with the physical qubits 3 and 4 form the maximally entangled state $|\Phi\rangle\langle\Phi|$ given by (17). We introduce another four qubits 5 –8 to describe the phase and the amplitude relaxations. These qubits interact with the physical qubits as described by the Hamiltonian in Section 3. Qubit 5 serves for the amplitude relaxation of the qubit 3 with the time-parameter $T_1$. Similarly, qubit 6 supports (?) the phase relaxation of the same qubit 3 with the time-parameter $T_2^{pure}$. In the same manner, qubits 7 and 8 support qubit 4.

In the Markovian approach, evolution is defined by the unitary operator in Hilbert space of all eight qubits

$$U = \exp\left(-iH\Delta t\right). \qquad (71)$$

At the beginning the state of the entire system is

$$\rho(0) = |\Phi\rangle\langle\Phi| \otimes |0\rangle\langle 0|. \qquad (72)$$

Similarly, the initial state before every evolution step is

$$\rho(t) = \rho_\chi(t) \otimes |0\rangle\langle 0|. \qquad (73)$$

Here $\rho_\chi(t)$ is the density matrix of the first four qubits (the chi-matrix at the time $t$ ).
The final state after every step is

$$\rho(t+\Delta t) = U\rho(t)U^\dagger. \qquad (74)$$

The last action of every step is Markovian reduction. This reduction is performed by tracing over the environment degrees of freedom and "forgetting" the system-environment correlations.

$$\rho(t+\Delta t) \to \rho_\chi(t+\Delta t) \otimes |0\rangle\langle 0|. \qquad (75)$$

We choose $\Delta t = (10^{-2} - 10^{-4})t_{oper}$ as the magnitude of the discretization time interval, where $t_{oper}$ is the total time of the operation. The results of modeling show that calculated values become independent of $\Delta t$ as $\Delta t$ decreases (for example, the results for $\Delta t = 10^{-2}t_{oper}$ are practically equal to the results for $\Delta t = 10^{-4}t_{oper}$ for the majority of cases).
In the following examples, we shall consider two types of the quantum noise: amplitude and phase relaxation introduced in Section 3, as well as the depolarizing noise.

The depolarizing noise is one that changes $s \times s$ density matrix as follows:

$$\rho \to \frac{pI}{s} + (1-p)U\rho U^\dagger. \qquad (76)$$

Here $I$ is the $s \times s$ unit matrix. The initial state is replaced by a fully chaotic state with probability $p$ , and we have an ideal unitary transform $U$ with probability $1-p$ .
Chi-matrix and Kraus operators of the depolarizing quantum noise can be constructed as follows. Let $e_U$ be the $s^2$ -length column normalized by 1, which is constructed by stretching the $s \times s$ unitary matrix $U$ (the 2-nd column is written under the first one etc.) and multiplying by the normalization coefficient $1/\sqrt{s}$ ( $e_U^\dagger e_U = 1$ ).
Consider the projector

$P = I - e_U e_U^\dagger$ , where $I$ is the identity $s^2 \times s^2$ matrix.

The matrix $P$ has $s^2$ eigenvalues. One of them is zero, and the others are equal to 1. We take the eigenvectors corresponding to nonzero eigenvalues and combine them into $s^2 \times (s^2-1)$ matrix $U_P$ . This matrix defines an orthogonal complement to $e_U$ .

Let us combine the column $\sqrt{1 - \frac{(s^2-1)p}{s^2}}e_U$ and the matrix $\sqrt{\frac{p}{s^2}}U_P$ into the single matrix

$$e = \left[ \sqrt{1 - \frac{(s^2 - 1)p}{s^2}} e_U \quad \sqrt{\frac{p}{s^2}} U_P \right]. \tag{77}$$

Finally, we obtain the target chi-matrix normalized by 1

$$\chi = ee^+. \tag{78}$$

The Kraus operators can be derived from the chi-matrix as described in section 2.

The first example of modeling is a CNOT chi-matrix (in Pauli matrices representation) on Fig. 3. The ideal case without noise is presented on the upper figure (a). The amplitude and phase relaxation with parameters $T_1 = 5$ and $T_2 = 3$ is presented on the middle figure (b) (the operation time is 1). The Hamiltonian of the operation is

$$H_{CNOT} = \begin{pmatrix} 0 & 0 & 0 & 0 \\ 0 & 0 & 0 & 0 \\ 0 & 0 & \pi/2 & -\pi/2 \\ 0 & 0 & -\pi/2 & \pi/2 \end{pmatrix}. \tag{79}$$

Finally, the chi-matrix of the CNOT gate with depolarizing noise ( $p = 0.6$ ) is presented on the bottom figure (c). Note that imaginary parts in figures (a) and (c) are equal to zero.

(a) ideal gate

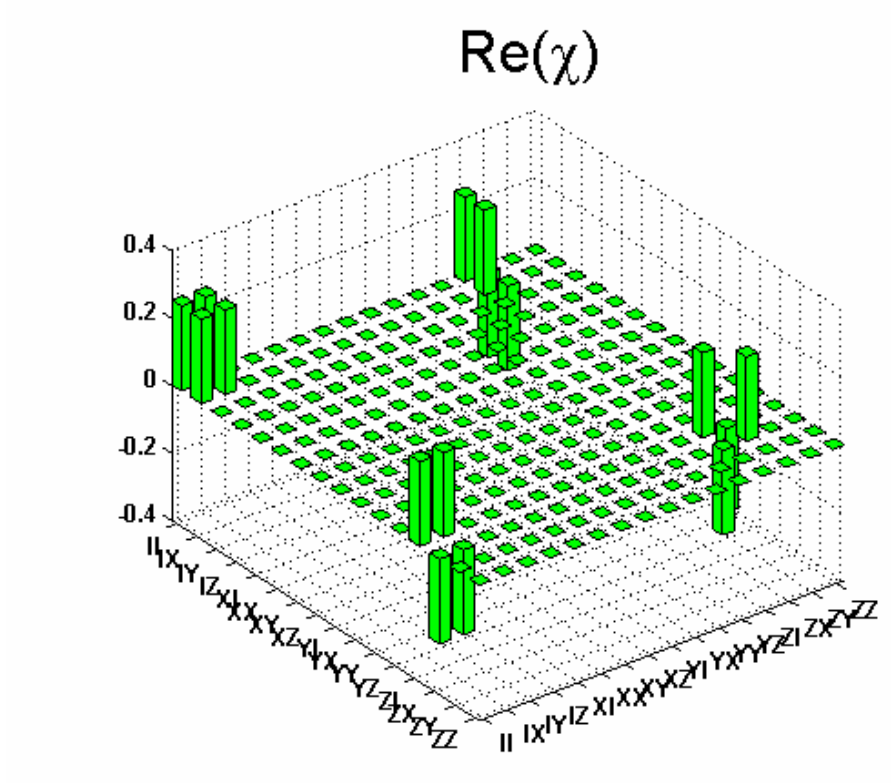

(b) amplitude and phase relaxation

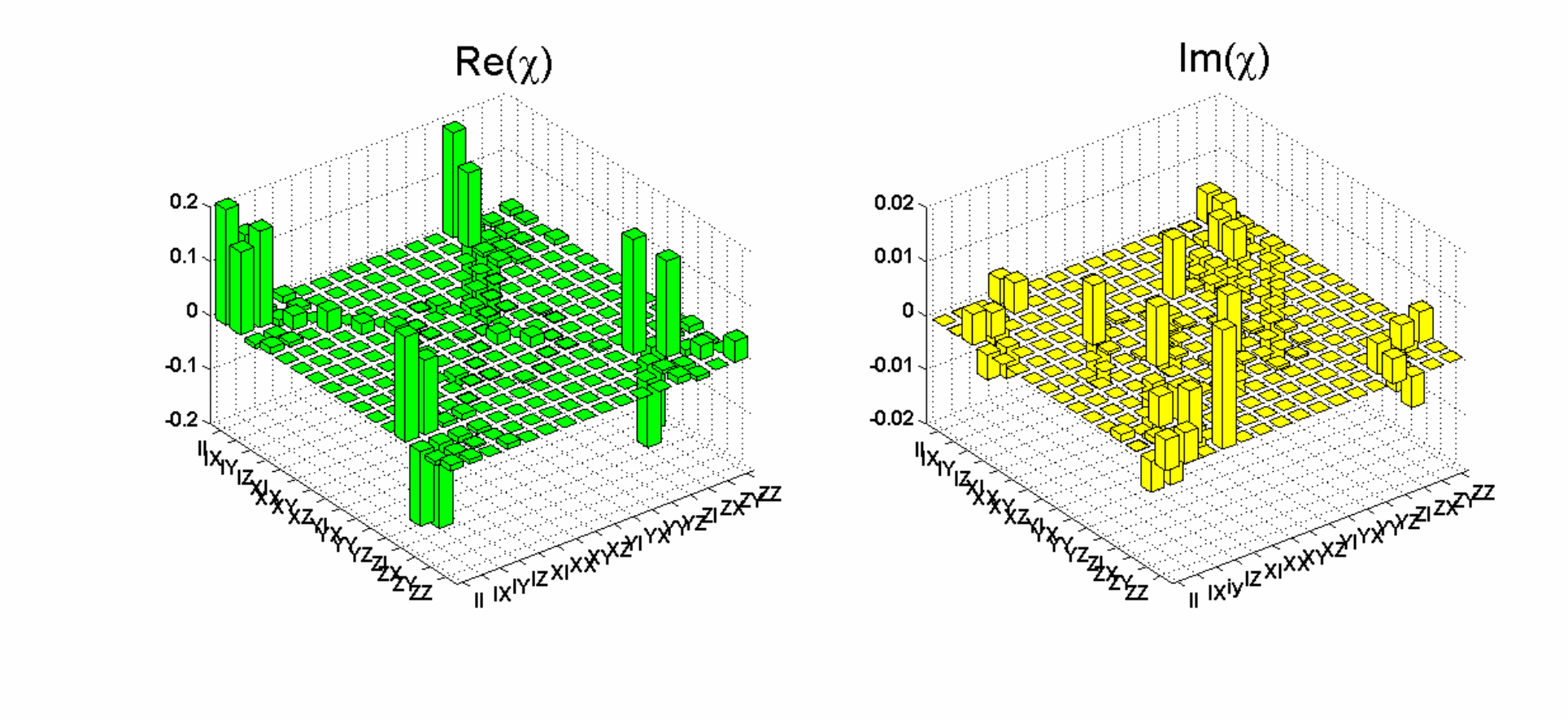

(c) depolarizing noise

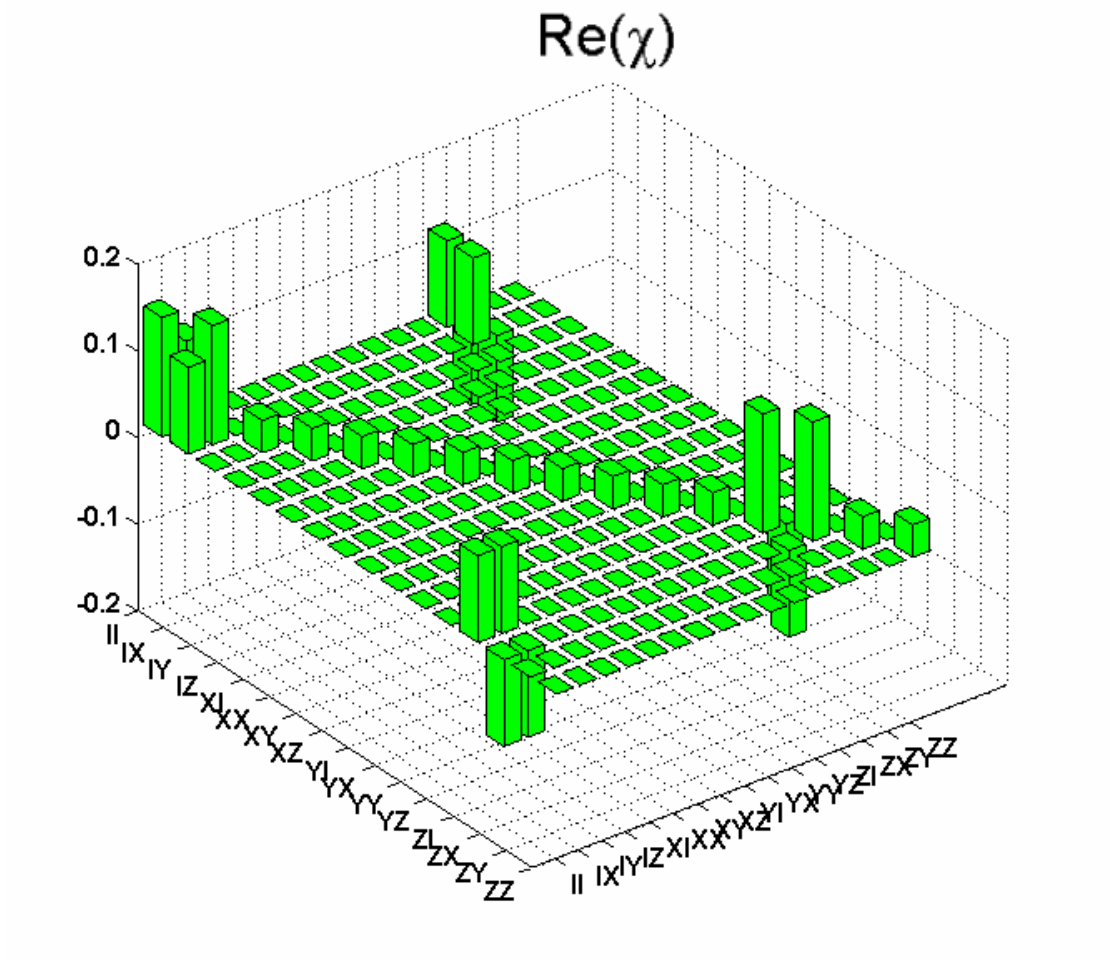

**Fig. 3.** Chi-matrix of the CNOT gate: (a) ideal gate without noise; (b) amplitude and phase relaxation with parameters $T_1 = 5$ and $T_2 = 3$ (the operation time is 1); (c) depolarizing noise ($p = 0.6$).

The next example is a noisy SQiSW gate. Here again, the ideal gate is presented on the figure (a), amplitude and phase relaxation on the figure (b), and, finally, the depolarizing noise ($p = 0.5$) on the figure (c).

A two-qubit SQiSW (square root of i-SWAP) transform appears as the result of the capacity (?) interaction between the superconducting qubits [21-23],

$$H_{\text{int}} = \hbar(g/2)\left(|01\rangle\langle 10| + |10\rangle\langle 01|\right), \tag{80}$$

where $g$ is the interaction constant.

Note that this Hamiltonian corresponds to the Heisenberg XY-interaction model.

Evolution of the system is determined by the unitary matrix

$$U_{\text{int}} = \begin{bmatrix} 1 & 0 & 0 & 0 \\ 0 & \cos(gt/2) & -i\sin(gt/2) & 0 \\ 0 & -i\sin(gt/2) & \cos(gt/2) & 0 \\ 0 & 0 & 0 & 1 \end{bmatrix}. \tag{81}$$

The pi-impulse ($gt = \pi$) performs the swap (i-swap): $|01\rangle \to -i|10\rangle$, $|10\rangle \to -i|01\rangle$. The half-length impulse ($gt = \pi/2$) performs the required SQiSW operation.

In the example illustrated by figure (b) the relaxation parameters are $T_1 = 3$ and $T_2 = 1.5$. Once again, we assume the time of operation to be equal to 1.

(a) ideal gate

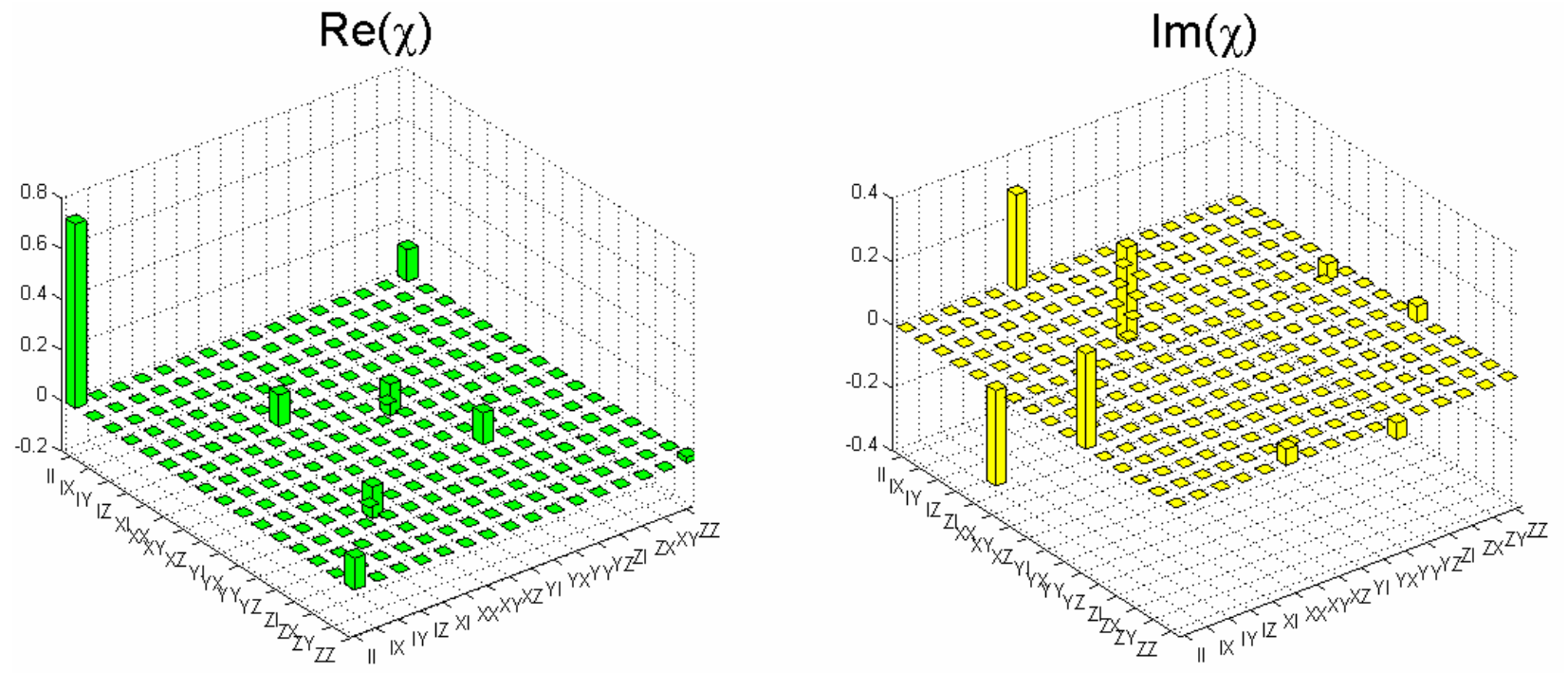

(b) amplitude and phase relaxation

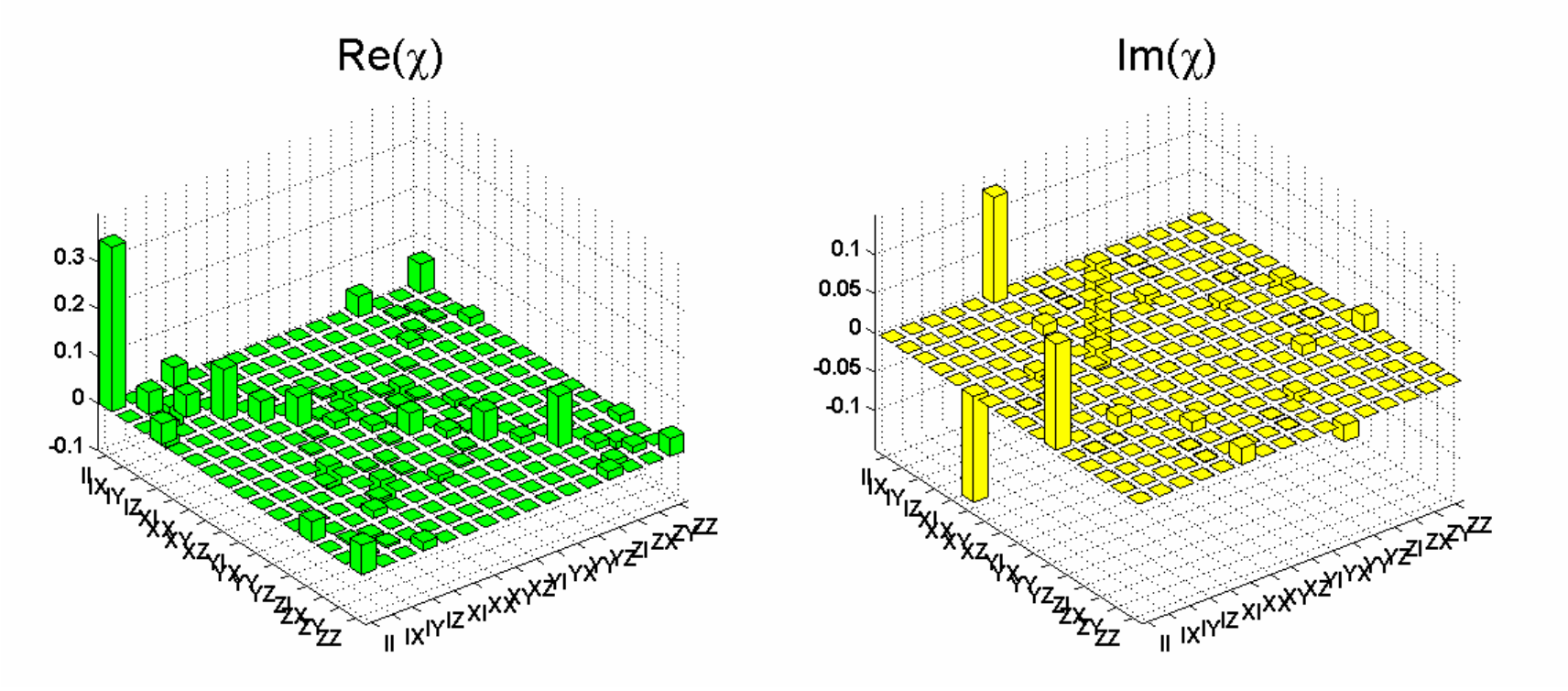

(c) depolarizing noise

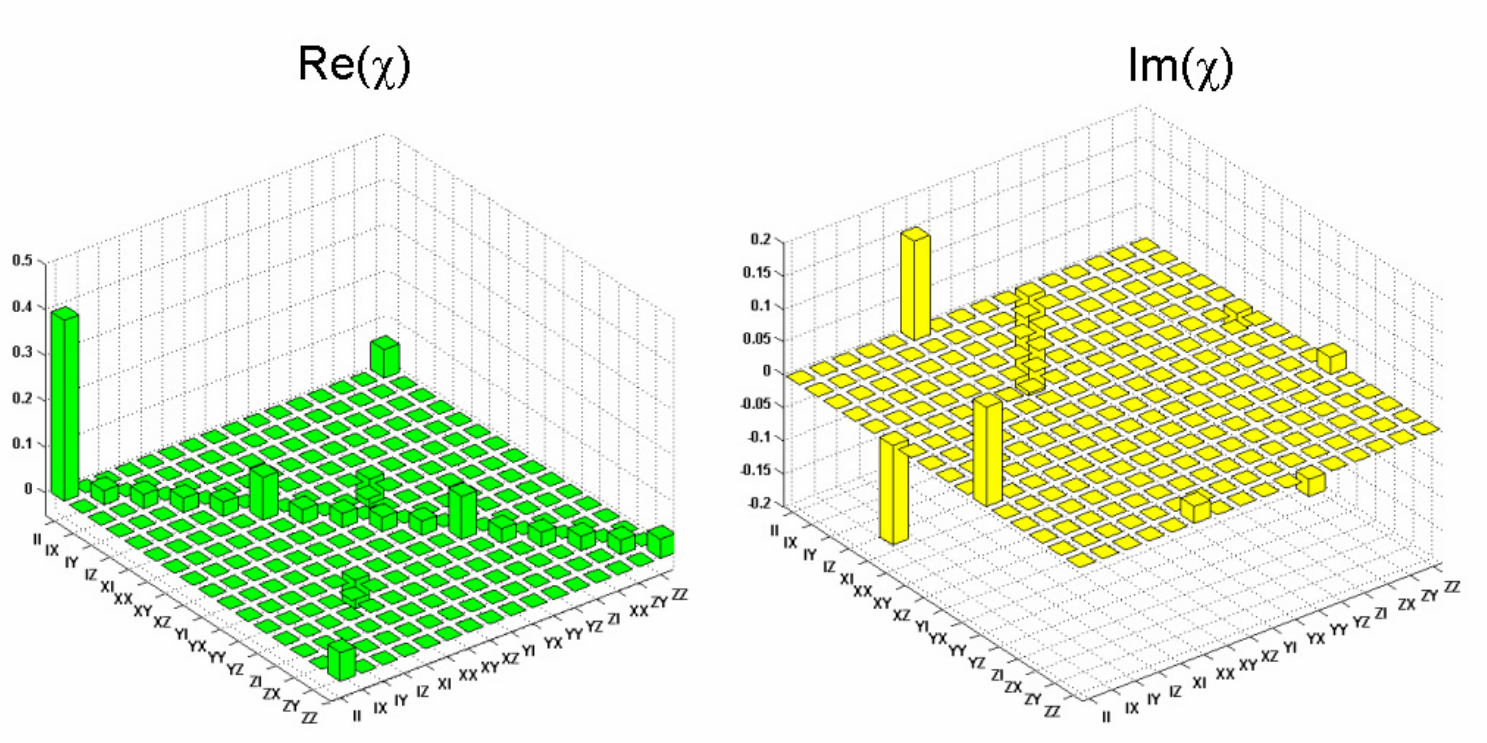

**Fig. 4.** Noise of the quantum SQiSW gate: (a) ideal gate; (b) chi-matrix with amplitude and phase relaxation; (c) depolarizing noise ( $p = 0.5$ ).

The next example considers modeling of a quantum CZ operation. Two-qubit CZ gate corresponds to the controlled Z-transformation: the controlled target-qubit undergoes the phase-flip transform $Z = \begin{pmatrix} 1 & 0 \\ 0 & -1 \end{pmatrix}$ if the controlling qubit is in the state $|1\rangle$. The corresponding unitary matrix is:

$$CZ = \begin{pmatrix} 1 & 0 & 0 & 0 \\ 0 & 1 & 0 & 0 \\ 0 & 0 & 1 & 0 \\ 0 & 0 & 0 & -1 \end{pmatrix}. \qquad (82)$$

This example illustrates one of the important practical features of the quantum state engineering: the use of ancillary quantum levels.

Consider a three-level quantum system consisting of a qubit and a qutrit (three-level quantum system). Once again, let the resonance interaction between $|11\rangle$ and $|02\rangle$ have the form defined by the Heisenberg XY-interaction:

$$H_{\text{int}} = \hbar (g/2)\left(|11\rangle\langle 02| + |02\rangle\langle 11|\right), \qquad (83)$$

where $g$ is the interaction constant.

The Hamiltonian in the matrix form is:

$$H_{\text{int}} = \begin{bmatrix} 0 & 0 & 0 & 0 & 0 & 0 \\ 0 & 0 & 0 & 0 & 0 & 0 \\ 0 & 0 & 0 & 0 & g/2 & 0 \\ 0 & 0 & 0 & 0 & 0 & 0 \\ 0 & 0 & g/2 & 0 & 0 & 0 \\ 0 & 0 & 0 & 0 & 0 & 0 \end{bmatrix}.$$

The unitary evolution is defined by the higher-dimensional version of (81):

$$U = \begin{bmatrix} 1 & 0 & 0 & 0 & 0 & 0 \\ 0 & 1 & 0 & 0 & 0 & 0 \\ 0 & 0 & \cos(gt/2) & 0 & -i\sin(gt/2) & 0 \\ 0 & 0 & 0 & 1 & 0 & 0 \\ 0 & 0 & -i\sin(gt/2) & 0 & \cos(gt/2) & 0 \\ 0 & 0 & 0 & 0 & 0 & 1 \end{bmatrix}. \quad (84)$$

The condition on time $gt = 2\pi$ implies the target CZ evolution

$$U(gt = 2\pi) = \begin{bmatrix} 1 & 0 & 0 & 0 & 0 & 0 \\ 0 & 1 & 0 & 0 & 0 & 0 \\ 0 & 0 & -1 & 0 & 0 & 0 \\ 0 & 0 & 0 & 1 & 0 & 0 \\ 0 & 0 & 0 & 0 & -1 & 0 \\ 0 & 0 & 0 & 0 & 0 & 1 \end{bmatrix}. \quad (85)$$

If we omit the 3-d and the 6-th columns and rows from the obtained $6 \times 6$ matrix, we will obtain the $4 \times 4$ matrix of CZ. Notice that the deleted rows and columns correspond to the $|02\rangle$ and $|12\rangle$ states, which are connected to the third level $|2\rangle$ of the qutrit. This level is necessary to construct the CZ gate, but we need the population of this level to be small (it must be zero in the perfect case). This ideal case is possible only without quantum noise and is practically unattainable due to the amplitude and phase relaxation.

We will describe the relaxation of the considered system similarly to Section 3. However, now the creation and annihilation operators will have $3 \times 3$ size. Actually, they will be the submatrices of the standard infinite matrices of creation and annihilation of the harmonic oscillator, namely

$$a = \begin{bmatrix} 0 & 1 & 0 \\ 0 & 0 & \sqrt{2} \\ 0 & 0 & 0 \end{bmatrix}, \qquad a^+ = \begin{bmatrix} 0 & 0 & 0 \\ 1 & 0 & 0 \\ 0 & \sqrt{2} & 0 \end{bmatrix}. \quad (86)$$

Another feature of this approach is the necessity to reduce the three-level system to a two-level system. The corresponding process is described by the following Kraus operators:

$$E_0 = \begin{bmatrix} 1 & 0 & 0 \\ 0 & 1 & 0 \end{bmatrix}, \; E_1 = \begin{bmatrix} 0 & 0 & 0 \\ 0 & 0 & 1 \end{bmatrix}. \quad (87)$$

The quantum operation is reduced to the identification of the level $|2\rangle$ with the level $|1\rangle$: $|0\rangle \to |0\rangle \; |1\rangle \to |1\rangle \; |2\rangle \to |1\rangle$.

In such an approach, the double perturbation is combined into a single one. It makes sense mostly for rough measurements (we can distinguish the unperturbed state from the perturbed one, but we can not split apart the one-photonic and two-photonic perturbations). For example, such situations often take place in the superconductive and optical technologies. Note also that the considered operation is trace-preserving:

$$E_0^+ E_0 + E_1^+ E_1 = \begin{bmatrix} 1 & 0 & 0 \\ 0 & 1 & 0 \\ 0 & 0 & 1 \end{bmatrix}. \quad (88)$$

An example of the CZ chi-matrix subject to relaxation is presented on Fig. 5b. The relaxation times for the first ( $A$ ) and second ( $B$ ) qubits are:

$$T_1^A = 3\ ,\ T_1^B = 4\ ,\ T_2^A = 2.4\ ,\ T_2^B = 3.2\ .$$

Once again, the (a) and (b) figures correspond to the ideal and depolarized gates ( $p = 0.4$ ).

(a) ideal gate

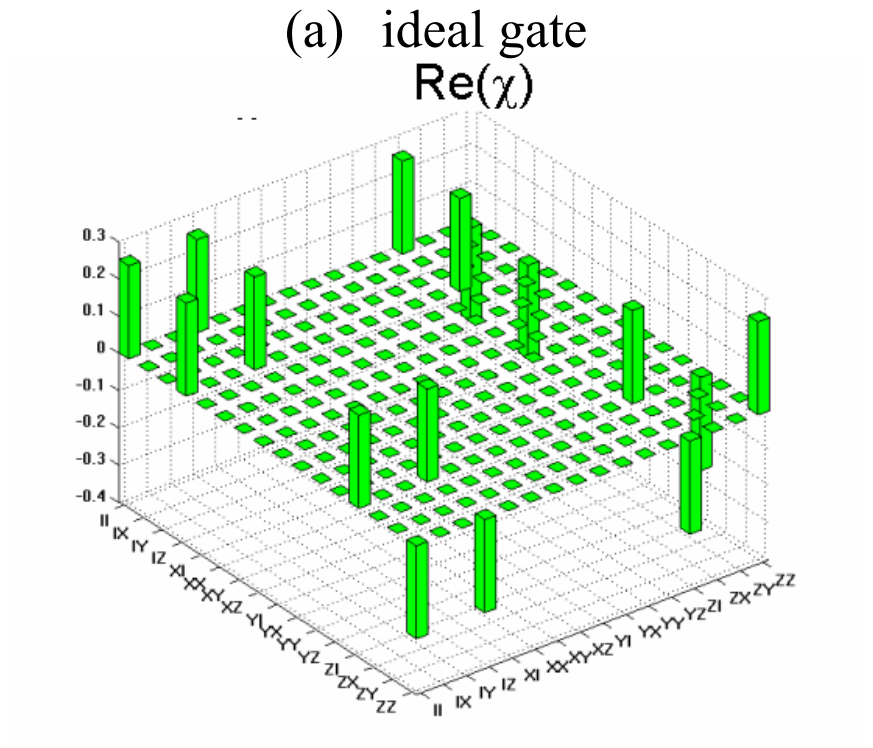

(b) amplitude and phase relaxation

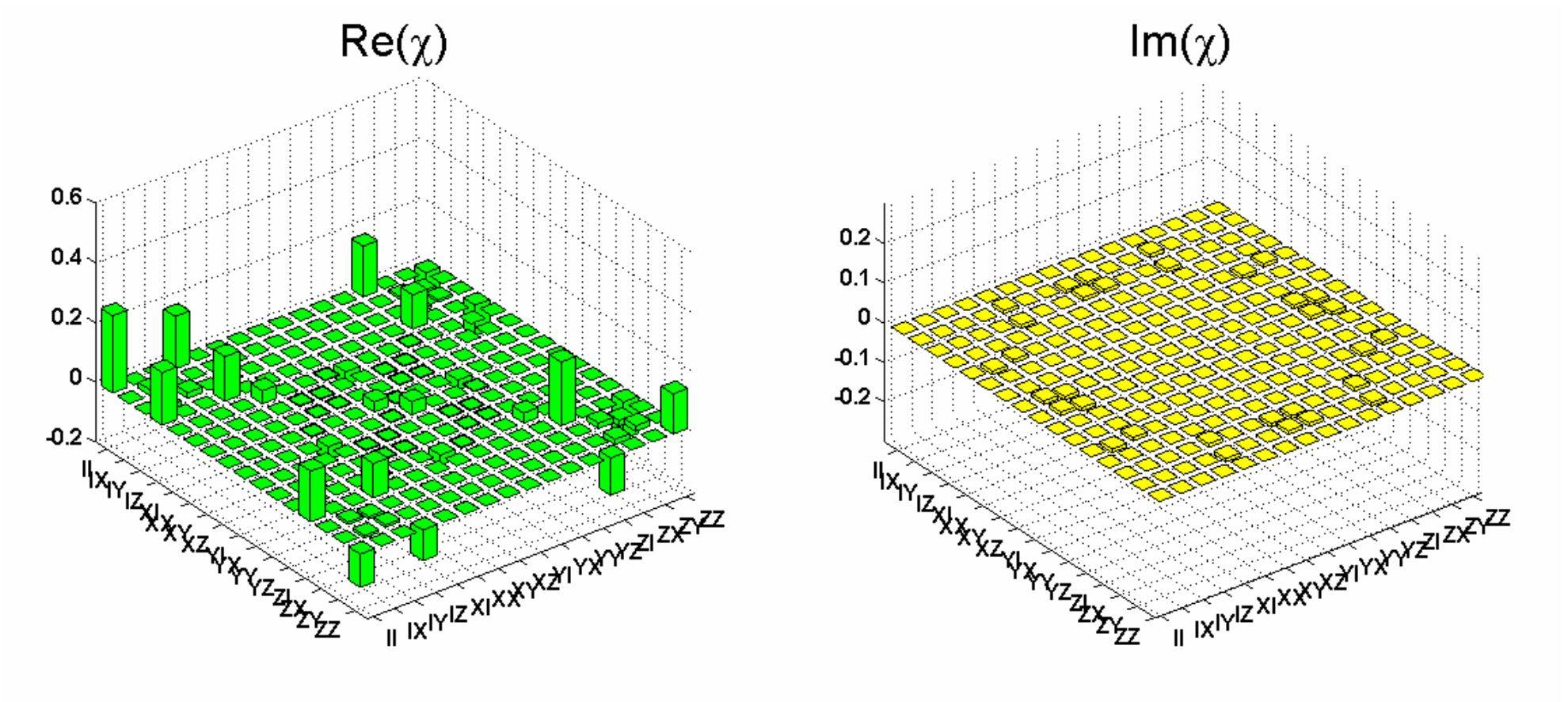

(c) depolarizing noise

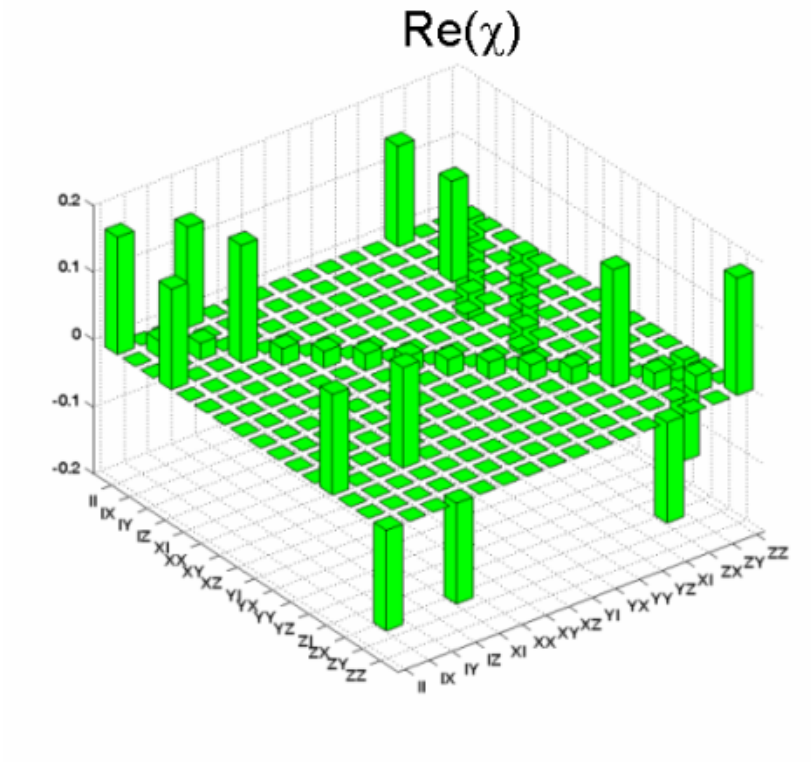

**Fig. 5**. Chi-matrix of CZ gate: (a) ideal case; (b) amplitude and phase relaxation; (c) depolarizing noise with $p = 0.4$.

The Choi-Jamiolkowski states framework may be successfully applied for the analysis of the quantum error correction codes.

For example, consider a standard three-qubit error correction scheme (Fig.6) that corrects the phase error (phase flip) and uses CNOT and Hadamard gates [3]. Notice that the phase error effect is equivalent to the phase-flip Z operator acting independently on every qubit with some probability $p$.

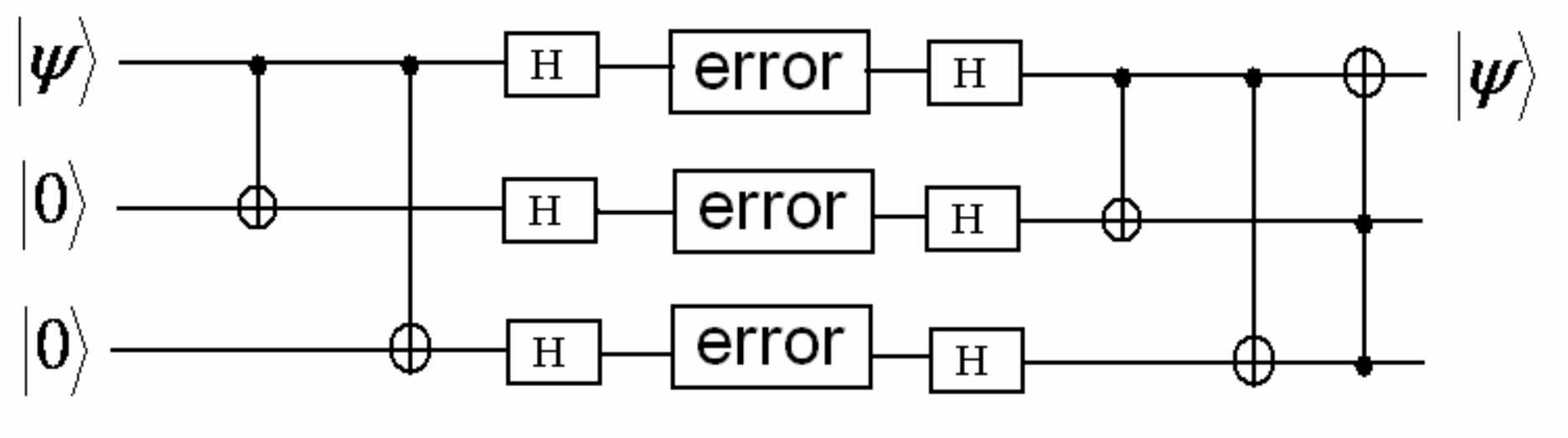

**Fig.6.** Quantum scheme of phase error correction algorithm

Let $\chi_0$ be the $4\times 4$ chi-matrix that corresponds to the ideal transformation (the identity operation in our case), while $\chi$ is the chi-matrix of the noisy gate with or without error correction. The degree of coincidence between the ideal and noisy chi-matrices can be described by the fidelity:

$$F = \left(Tr\sqrt{\chi_0^{1/2}\chi\chi_0^{1/2}}\right)^2 . \tag{89}$$

We assume that the matrices are trace-normalized to one in this case.

The chi-matrices of ideal and noisy (with probability p) gates are, respectively,

$$\chi^{Ideal} = \frac{1}{2}\begin{pmatrix} 1 & 0 & 0 & 1 \\ 0 & 0 & 0 & 0 \\ 0 & 0 & 0 & 0 \\ 1 & 0 & 0 & 1 \end{pmatrix}, \qquad \chi^{Noise} = \frac{1}{2}\begin{pmatrix} 1 & 0 & 0 & 1-2p \\ 0 & 0 & 0 & 0 \\ 0 & 0 & 0 & 0 \\ 1-2p & 0 & 0 & 1 \end{pmatrix} \tag{90}$$

Note that the scheme on Fig.6 shows the one-qubit transformation and not a three-qubit operation as it may seem at first glance. However, the inputs of the two ancilla qubits are fixed and the chi-matrix of this operation is

$$\chi^{Code} = \frac{1}{2}\begin{pmatrix} 1-3p^2+2p^3 & 0 & 0 & 1-3p^2+2p^3 \\ 0 & 3p^2-2p^3 & 3p^2-2p^3 & 0 \\ 0 & 3p^2-2p^3 & 3p^2-2p^3 & 0 \\ 1-3p^2+2p^3 & 0 & 0 & 1-3p^2+2p^3 \end{pmatrix}. \tag{91}$$

We can represent the considered chi-matrices in the Bell-states basis:

$$\chi^{Ideal} = \left|\Phi^+\right\rangle\left\langle\Phi^+\right|, \tag{92}$$

$$\chi^{Code} = \left(1-3p^2+2p^3\right)\left|\Phi^+\right\rangle\left\langle\Phi^+\right| + \left(3p^2-2p^3\right)\left|\Psi^+\right\rangle\left\langle\Psi^+\right|, \tag{93}$$

$$\chi^{Noise} = \left(1-p\right)\left|\Phi^+\right\rangle\left\langle\Phi^+\right| + p\left|\Phi^-\right\rangle\left\langle\Phi^-\right|, \tag{94}$$

where

$$\left|\Phi^+\right\rangle = \frac{1}{\sqrt{2}}\left(|00\rangle + |11\rangle\right) = \frac{1}{\sqrt{2}}\begin{pmatrix} 1 \\ 0 \\ 0 \\ 1 \end{pmatrix}, \quad \left|\Psi^+\right\rangle = \frac{1}{\sqrt{2}}\left(|01\rangle + |10\rangle\right) = \frac{1}{\sqrt{2}}\begin{pmatrix} 0 \\ 1 \\ 1 \\ 0 \end{pmatrix},$$

$$\left|\Phi^-\right\rangle = \frac{1}{\sqrt{2}}\left(|00\rangle - |11\rangle\right) = \frac{1}{\sqrt{2}}\begin{pmatrix} 1 \\ 0 \\ 0 \\ -1 \end{pmatrix}. \tag{95}$$

From the chi-matrix expressions introduced above we can get the desired fidelity values:

$$F\left(\chi^{Ideal}, \chi^{Noise}\right) = 1 - p, \tag{96}$$

$$F\left(\chi^{Ideal}, \chi^{Code}\right) = 1 - 3p^2 + 2p^3. \tag{97}$$

The derived result is presented on Fig. 7

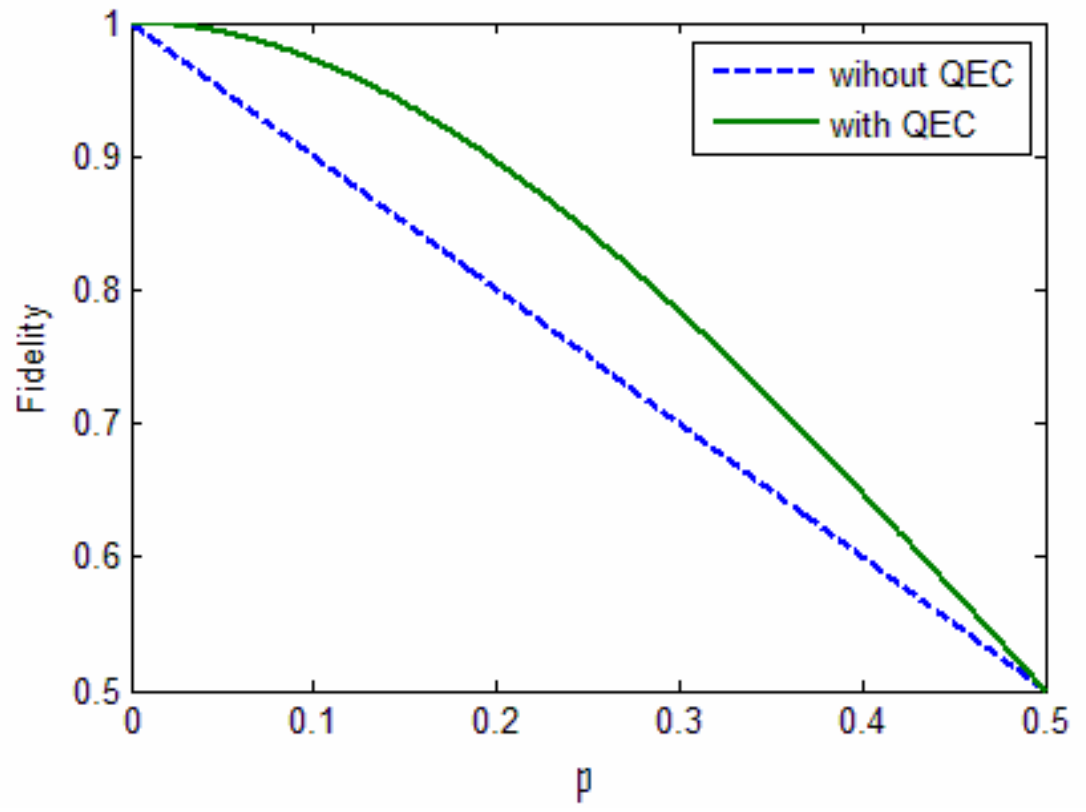

**Fig. 7.** Illustration of the use of the Choi-Jamiolkowski formalism for analysis of the phase-flip correction code

We can see that for small $p$*'s* the correction code leads to a slower quadratic dependence instead of the linear one. These results coincide with the results of the well-known approach ([3]?) but differ from a methodological point of view. Our approach is based on the concept of fidelity (89) of two quantum operations. Here, the Choi-Jamiolkoski state represents a quantum operation as a whole. In our opinion, the proposed approach is simpler than the traditional one based on the minimal fidelity for **individual states** [3], because in the latter case we need to consider all possible input states and solve the optimization problem.

We have considered here only the problem of phase flip error correction. The quantum error detection, which is associated with the energy relaxation, is discussed in the recent paper [24].

## 5. Dynamics of the quantum operations entanglement

The Choi-Jamiolkowski state characterized by a chi-matrix contains complete information about the corresponding quantum operation. One may expect that the study of entanglement characteristics of the Choi-Jamiolkowski state could provide an important information about the emergence, maintaining and breaking of the entanglement in the quantum operations. Consider a two-qubit quantum operation and the corresponding four-qubit chi-matrix. Here again, we assume that qubits 1 and 2 form the ancillary system and qubits 3 and 4 are the components of the open physical system. Consider two ways of splitting the joint system into subsystems.

In the first approach qubits 1 and 2 form the first subsystem while qubits 3 and 4 form the second subsystem. In the second approach the first subsystem is formed by qubits 1 and 3 and the second one is formed by qubits 2 and 4.Our goal is to study entanglement between the corresponding subsystems.

In the first case subsystems $A$ and $B$ are initially in the maximally entangled pure state $|\Phi\rangle$ described by (17). The decoherence process due to quantum noise leads to a gradual break-up of the entanglement (in other words, the channel is subject to partial or full loss of the entanglement).

The situation of spatial separation of the subsystems $A$ and $B$ is of particular interest to the quantum information technology. In this case, entanglement between the distant subsystems allows one to perform such specific tasks as quantum teleportation, quantum dense coding, quantum cryptography protocols etc.

The second opposite case is that of initially non-entangled subsystems $A'$ and $B'$. Then our study concerns the important property of two- and multi-qubit gates to produce and maintain the entanglement resource even under the quantum noise. The entanglement that emerges during the evolution process can be treated as the correlation between two quantum channels linked to the physical qubits.

We shall use the so-called negativity measure to estimate the amount of entanglement between the subsystems. This is an easy-to-compute measure of entanglement, equal to the absolute value of the sum of all negative eigenvalues of the partially transposed density matrix. We can compute the negativity by the formula:

$$Negativity = \frac{1}{2}\left( \sum_j \left|\lambda_j^{pt}\right| - \sum_j \lambda_j^{pt} \right). \qquad (98)$$

In our case the chi-matrix $\chi$ has unit trace and plays the role of a density matrix. In the equation above $\lambda_j$ are the eigenvalues of $\chi$; $\lambda_j^{pt}$ are the eigenvalues of partially-transposed chi-matrix $\chi^{pt}$,. This matrix is generated by the transform $T_A \otimes E_B$ which transposes subsystem $A$ and leaves unchanged subsystem $B$. Generally speaking, this transform is not physical, because it can lead to a nonpositive output matrix if the subsystems are entangled. Thus, the presence of the negative eigenvalues in the matrix $\chi^{pt}$ is a necessary condition for inseparability of the subsystems $A$ and $B$ . Note that this condition is also sufficient for subsystems of $2 \times 2$ , $2 \times 3$ or $3 \times 2$ dimensions (the so-called Peres–Horodeckis criterion).

In the case of two-qubit operations both $\chi$ and $\chi^{pt}$ matrices are of $16 \times 16$ size. An explicit equation connecting the matrix elements for the first-type split (into the subsystems $A = \{1,2\}$ and $B = \{3,4\}$) is:

$$\chi^{\mathrm{T}^{12}}_{8k_1+4k_2+2j_3+j_4+1,8j_1+4j_2+2k_3+k_4+1} = \chi_{8j_1+4j_2+2j_3+j_4+1,8k_1+4k_2+2k_3+k_4+1}, \qquad (99)$$

$$j_1, j_2, j_3, j_4, k_1, k_2, k_3, k_4 = 0,1.$$

A similar formula for the second-type split (into the subsystems $A' = \{1,3\}$ and $B' = \{2,4\}$) is

$$\chi^{\mathrm{T}^{13}}_{8k_1+4j_2+2k_3+j_4+1,8j_1+4k_2+2j_3+k_4+1} = \chi_{8j_1+4j_2+2j_3+j_4+1,8k_1+4k_2+2k_3+k_4+1}, \qquad (100)$$

$$j_1, j_2, j_3, j_4, k_1, k_2, k_3, k_4 = 0,1.$$

The equations presented above can be generalized to the arbitrary dimensions. For example, if a quantum operation acts in $s$-dimensional Hilbert space, the first-type split is described by :

$$\chi^{\mathrm{T}^{\mathrm{A}}}_{s(k_1-1)+j_2,s(j_1-1)+k_2} = \chi_{s(j_1-1)+j_2,s(k_1-1)+k_2}, \qquad (101)$$

$$j_1, j_2, k_1, k_2, = 1,\ldots,s.$$

Note that zero negativity defines the so-called PPT-channels [6]. The Choi-Jamiolkowski states of these channels remain positive under a partial transpose. Entanglement-breaking channels are an important particular case of PPT-channels. In general, PPT-states have an entanglement, but this entanglement is in some sense “weak” because it does not allow for distillation.

Fig.8 illustrates breaking of entanglement of the initially maximally entangled state under the depolarizing noise. Here the dimensions of the states vary from $s = 2$ to $s = 10$.

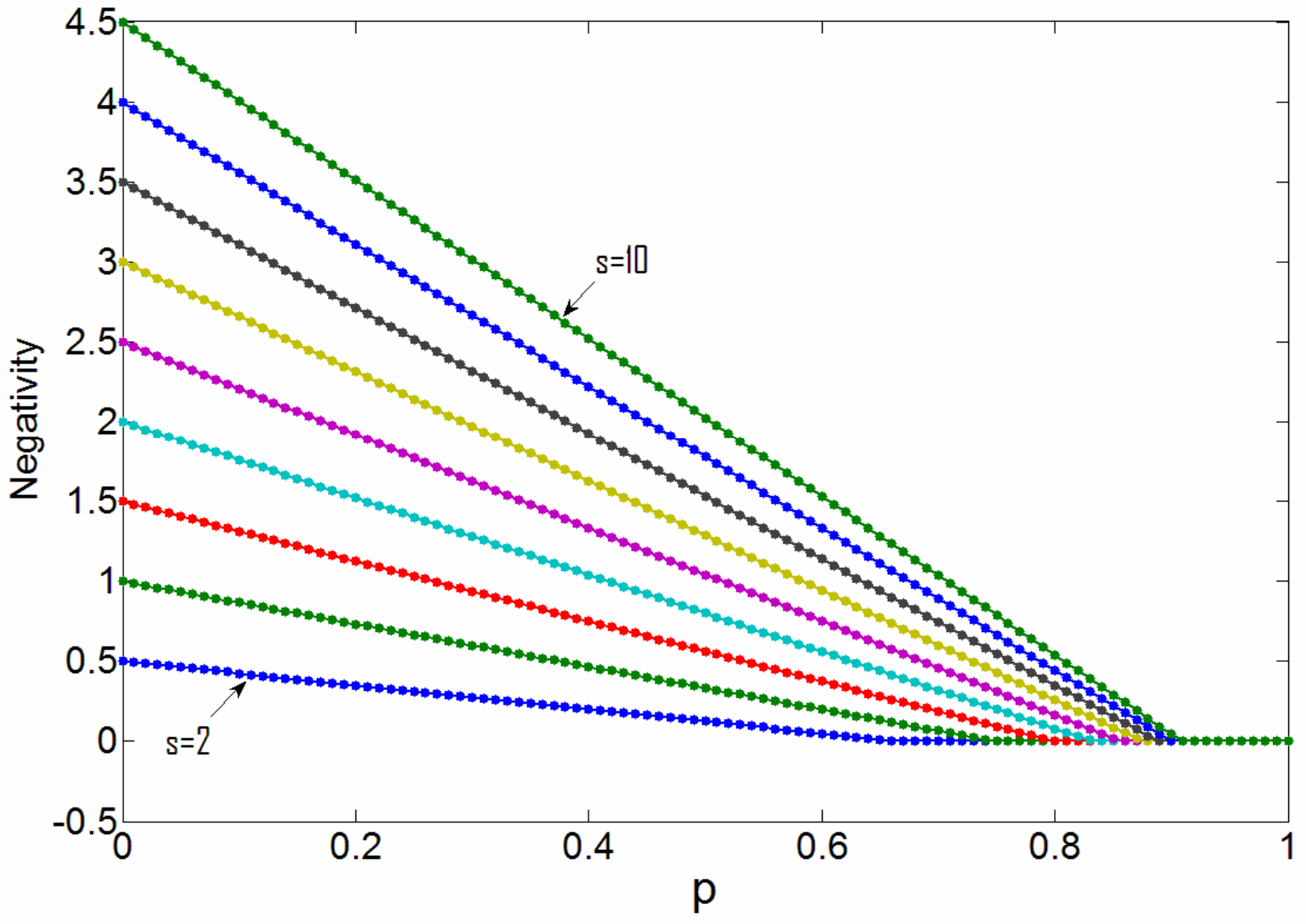

**Fig. 8.** Breaking of entanglement under the depolarizing noise in the Hilbert spaces of dimensions from $s=2$ to $s=10$ (from bottom to top).

Notice that negativity decreases linearly from the maximum (no noise) to zero for the critical noise value

$$p_c = \frac{s}{s+1}. \tag{102}$$

This value can be also derived theoretically [6,25].

The next Fig.9 illustrates the breaking of entanglement under the amplitude and phase relaxation for the case of two-qubit and four-qubit quantum operations ( $s=2$ and $s=4$ ). The parameters of the relaxation are $T_1 = 5$, $T_2 = 3$.

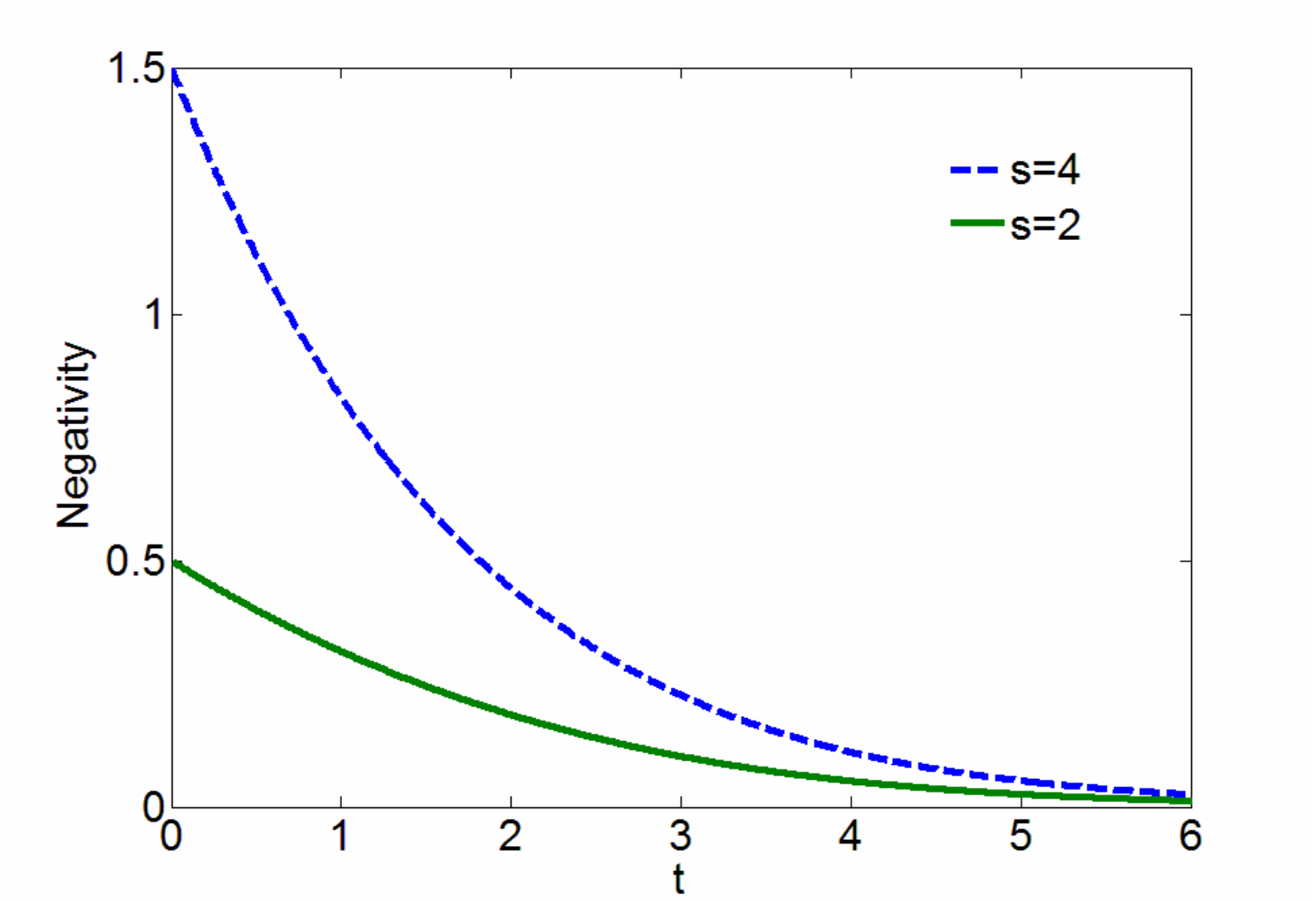

**Fig.9.** Breaking of entanglement under the amplitude and phase relaxation for the case of two-qubit and four-qubit quantum operations ( $s=2$ and $s=4$ ). The parameters of relaxation are $T_1 = 5$, $T_2 = 3$.

Note that for the first-type split negativity is independent of the operations acting on the system (i.e. single-, two- or multi-qubit operations), because in this case all operations are local.

Now consider emergence and breaking of the entanglement for the second-type split. Here the initial entanglement between the subsystems $A'$ and $B'$ is zero, and the results critically depend on the type of the quantum operation considered.

Fig. 10 illustrates the dynamics of the entanglement for the Heisenberg XY-interaction. The corresponding Hamiltonian is

$$H_{\text{int}} = \hbar(\pi/4)\left(|01\rangle\langle 10| + |10\rangle\langle 01|\right). \tag{103}$$

The SQiSW gate is realized for $t = 1$.

The ideal dynamics without noise is presented on the upper figure, and dynamics under the amplitude and phase relaxation with parameters $T_1 = 7$ $T_2 = 5$ on the lower figure.

a) The ideal gate

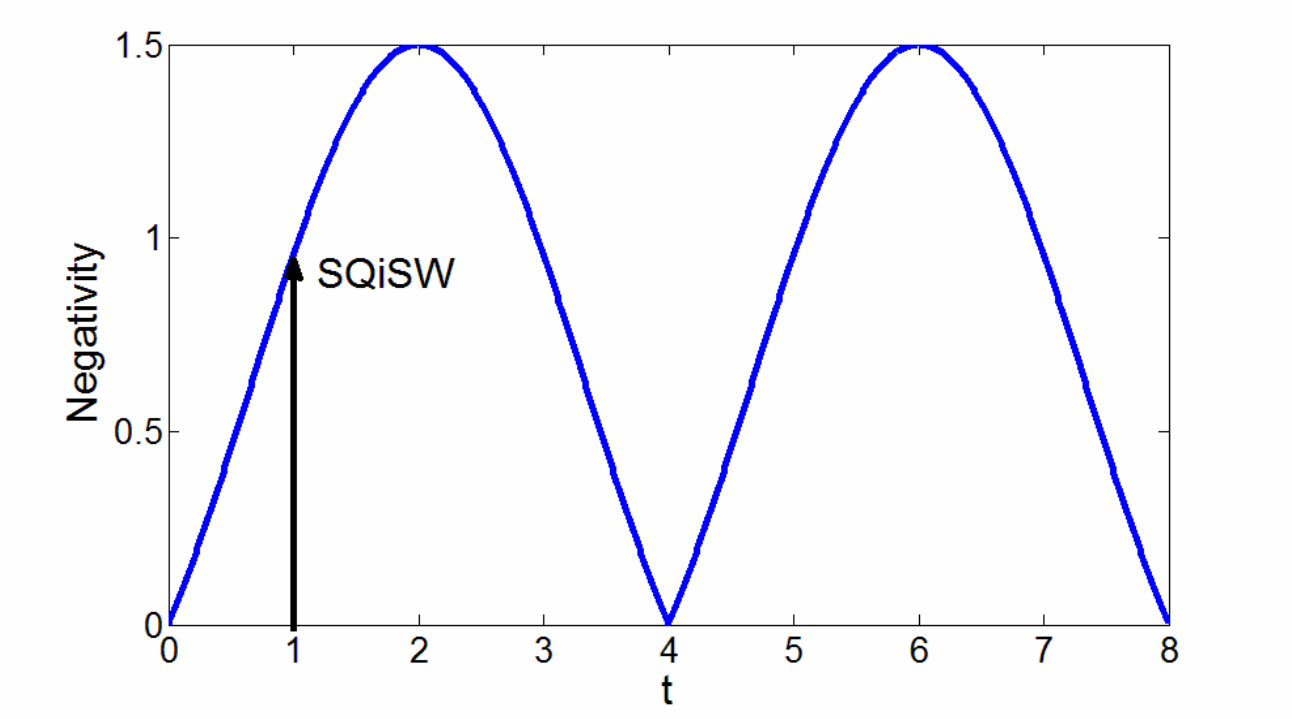

b) The gate under the amplitude and phase relaxation

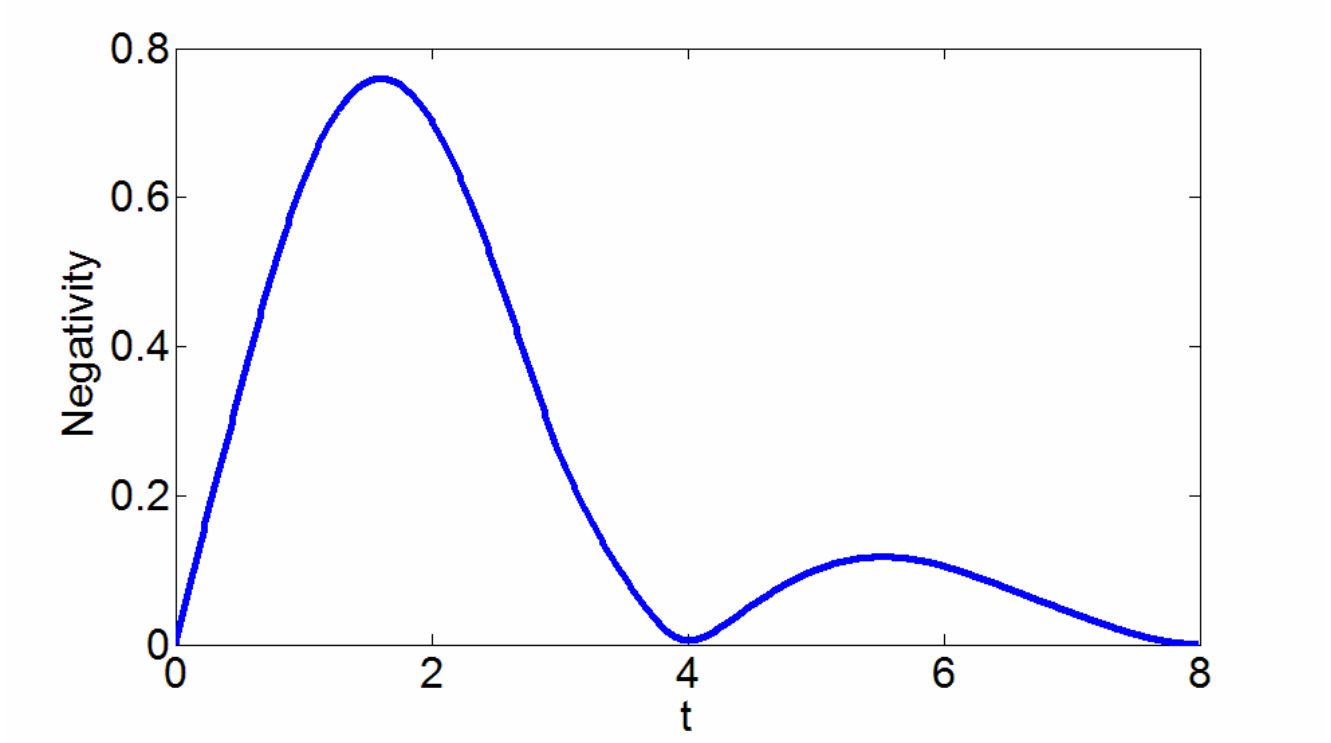

**Fig.10.** Dynamics of the entanglement for the Heisenberg XY-interaction: (a) ideal gate; (b) under the amplitude and phase relaxation

The next Fig.11 shows similar behavior for the CNOT gate.

a) The ideal gate

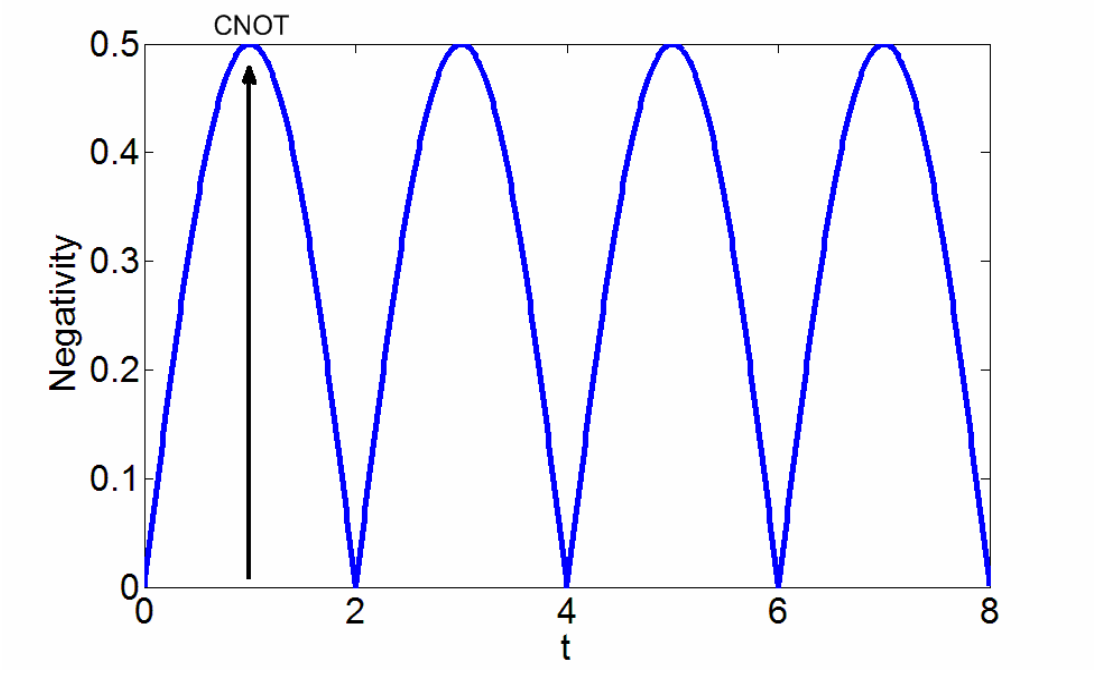

b) The gate under the amplitude and phase relaxation

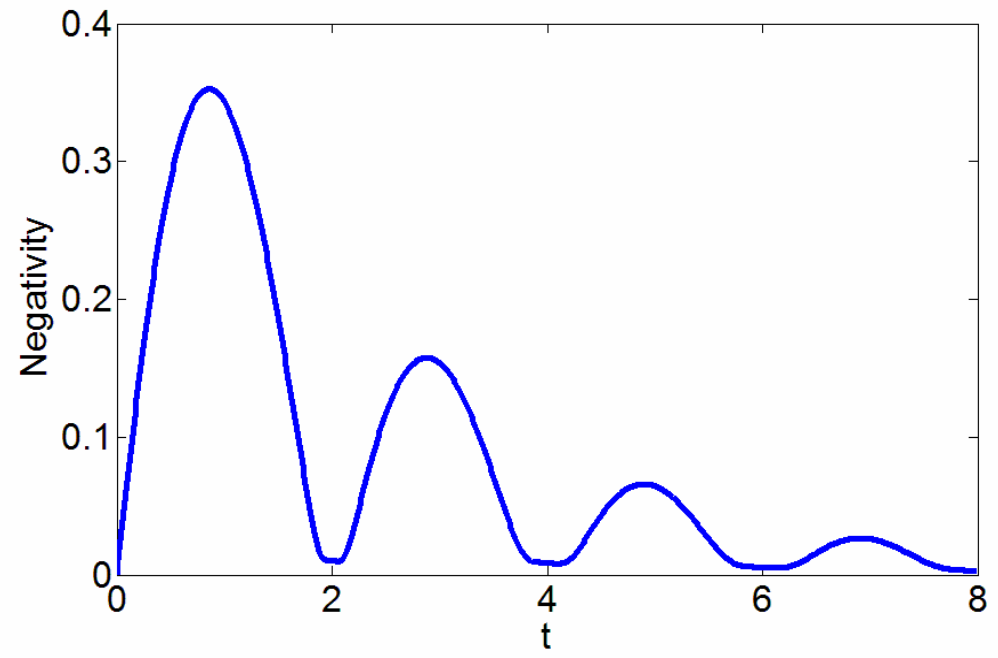

**Fig.11.** Dynamics of entanglement for the CNOT Hamiltonian: (a) ideal gate; (b) under the amplitude and phase relaxation

Figure 12 shows the entanglement of the $\widetilde{\chi}$ matrix, which describes "pure" noise. A relatively small residual entanglement that is observed in this case characterizes the correlations between the errors that appear in different qubits during the two-qubit gates realizations. This effect impedes implementation of quantum error correction algorithms, because the latter usually aim to correct separate independent errors.

The upper figure shows the residual entanglement for the Heisenberg XY-model, the lower graph illustrates the dynamics for the CNOT Hamiltonian. The parameters of relaxation are once again $T_1 = 7$ , $T_2 = 5$ .

a) The gate SQiSW

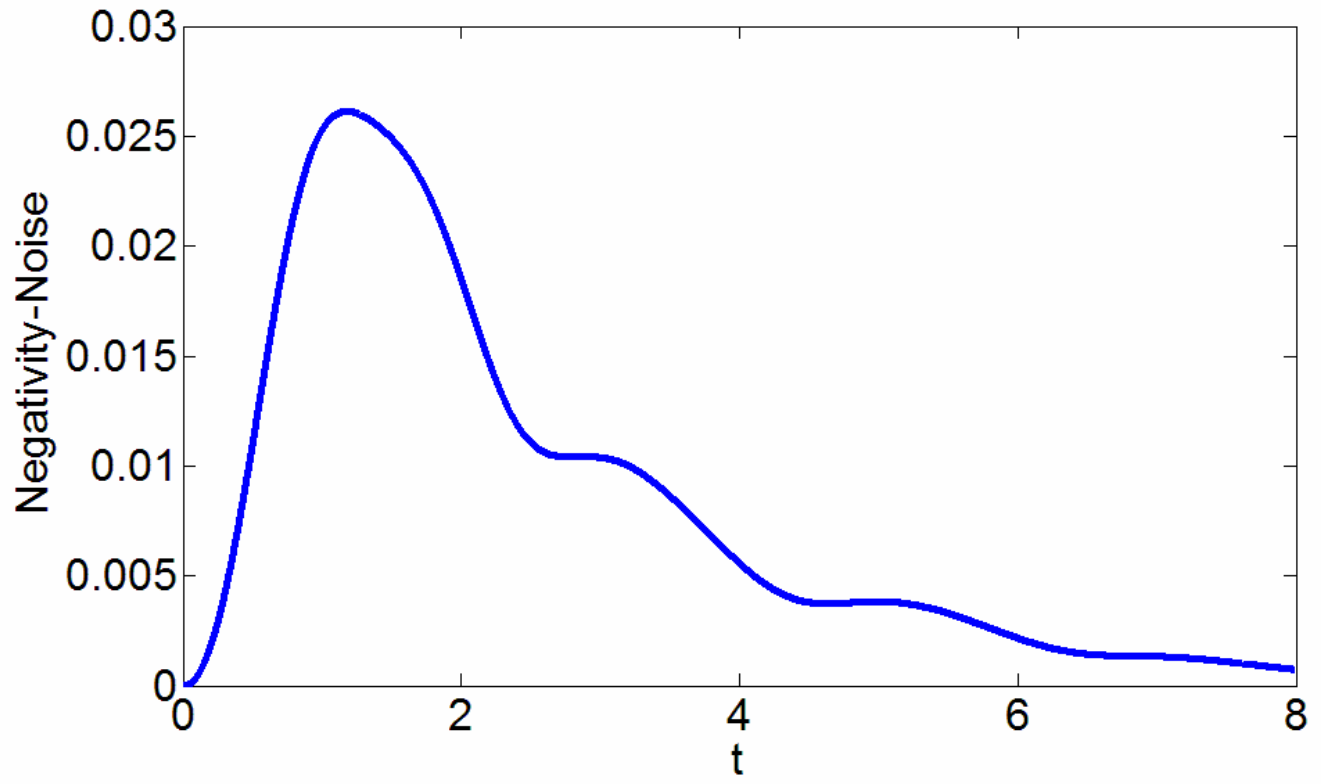

b) The gate CNOT

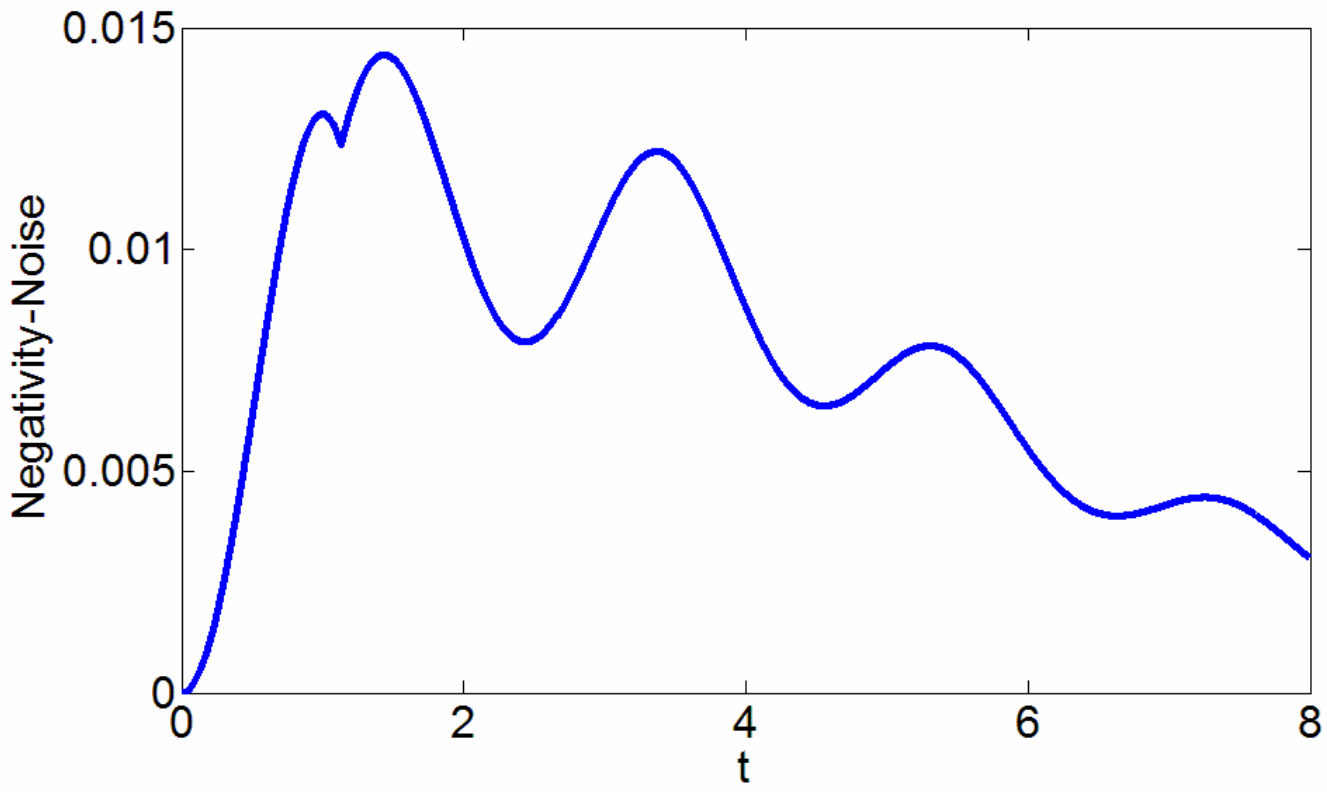

**Fig. 12.** Dynamics of the entanglement of quantum noise due to phase and amplitude relaxation for (a) $XY$ - interaction model, (b) CNOT Hamiltonian

## 6. Conclusions

Let us briefly formulate the main results of the paper.

1. Methods and algorithms of mathematical modeling of quantum operations were developed having in mind the tasks of design of the quantum computers hardware. We reviewed the basics of quantum operations theory emphasizing certain matrix representations of main objects, which appear quite useful for the algorithms implementation. Different ways of description of quantum processes were considered, including the operators-sum, the unitary dilation, the Choi-Jamiolkowski states (chi-matrix), and the evolution operator. The method of modeling the amplitude and phase relaxation acting simultaneously with the Hamiltonian evolution was described in detail.
2. Some practically important quantum gates, including SQiSW (square root of i-SWAP), controlled NOT (CNOT), and controlled Z-transform, were modeled taking into account different mechanisms of decoherence (depolarizing noise, amplitude and phase relaxation).
3. By the example of the phase-flip error correction, the problem of quantum error correction in the framework of the Choi-Jamiolkowski states was considered. This approach allows one to estimate a quantum operation **as a whole**. It is an alternative to the standard approach [3] which is based on the lower bounds for the fidelity between **individual states** and on the solution of the corresponding minimization problem.
4. The problem of emergence, maintaining and breaking of the entanglement for two- and multi-qubit quantum operations in quantum noise was considered. It was shown that different splitting of the Choi-Jamiolkowski state into subsystems leads to two practically important measures of entanglement.

We would like to thank Prof. A.N. Korotkov for helpful discussions.

This work was supported in part by the Fundamental Research Program of Russian Academy of Sciences.